\documentclass[12pt]{article}
\usepackage{amsthm, amsmath, amssymb, amsfonts,euscript,mathrsfs,mathabx}
\usepackage{bm,bbm,latexsym}
\usepackage{psfig, epsfig, verbatim}
\usepackage{natbib}
\usepackage{times}
\usepackage{setspace}
\usepackage{multirow}
\usepackage{color}
\usepackage{subfig}
\usepackage{float}
\usepackage{cleveref}

\newcommand\independent{\protect\mathpalette{\protect\independenT}{\perp}}
\def\independenT#1#2{\mathrel{\rlap{$#1#2$}\mkern2mu{#1#2}}}

\newcommand{\NOx}{NO$_x$}

\newcommand{\noteC}[1]{{\footnote{\color{red} \bf{Note:} #1} }}
\newcommand{\noteG}[1]{{\footnote{\color{blue} \bf{GP:} #1} }}

\newcommand{\noti}{(-i)}
\newcommand{\pnoti}{p_{k \ne i} \in T_j}
\newcommand{\pstar}{p_{i(j)}^*}

\newcommand{\setsizej}{|T_j|}
\newcommand{\setsizei}{|T^\top_i|}

\newcommand{\pclustsize}{|P^k|}
\newcommand{\mclustsize}{|M^k|}

\newcommand{\Aintsub}{\bA_{\{T_j=1\}}}
\newcommand{\Aintsubnoti}{\bA_{\{T_j\ne1\}}}

\newcommand{\intsetj}{T_j}
\newcommand{\intseti}{T^\top_i}

\newcommand{\denominator}{f_{\bA | \bW, \bX, k}(\bA^k | \bW^k, \bX^k)}
\newcommand{\numerator}{\pi( \bA^k_{(-i^*)}| A_{i^*}=a, \alpha)}
\newcommand{\numzips}{23,458 }

\newtheorem{assumption}{Assumption}

\title{Bipartite Causal Inference with Interference} 
\date{}
\author{Corwin M. Zigler\thanks{Harvard T.H. Chan School of Public Health} \and Georgia Papadogeorgou\thanks{Duke University}}

\begin{document}
\maketitle

\abstract{Statistical methods to evaluate the effectiveness of interventions are increasingly challenged by the inherent interconnectedness of units.  Specifically, a recent flurry of methods research has addressed the problem of {\it interference} between observations, which arises when one observational unit's outcome depends not only on its treatment but also the treatment assigned to other units.   We introduce the setting of {\it bipartite causal inference with interference,} which arises when 1) treatments are defined on observational units that are distinct from those at which outcomes are measured and 2) there is {\it interference} between units in the sense that outcomes for some units depend on the treatments assigned to many other units.  Basic definitions and formulations are provided for this setting, highlighting similarities and differences with more commonly considered settings of causal inference with interference.  Several types of causal estimands are discussed, and a simple inverse probability of treatment weighted estimator is developed for a subset of simplified estimands.  The estimators are deployed to evaluate how interventions to reduce air pollution from  473 power plants in the U.S. causally affect cardiovascular hospitalization among Medicare beneficiaries residing at \numzips zip code locations.}

\newcommand{\nc}{\newcommand}
\nc{\nn}{\nonumber}
\nc{\fns}{\footnotesize}

\nc{\revisionline}{\vspace{.1in} \today \vspace{.1in} \hrule\hrule\hrule\vspace{.1in}}
\nc{\newpp}{\vspace{.1in} \noindent}

\nc{\slideline}{\smallskip \hrule\hrule \smallskip}

\nc{\wh}{\widehat}

\def\boxit#1{\vbox{\hrule\hbox{\vrule\kern6pt
          \vbox{\kern6pt#1\kern6pt}\kern6pt\vrule}\hrule}}

\newcommand{\se}{\text{se} }
\newcommand{\spec}{\text{sp} }
\newcommand{\fpr}{\text{{\sc fpr}}}
\newcommand{\fnr}{\text{{\sc fnr}}}

\nc{\Ef}{ {\rm E}_{\infty} }
\nc{\Ex}{ {\rm E} }
\nc{\Ec}{ {\rm E}_1 }
\nc{\Pf}{ {\rm P}_{\infty} }
\nc{\Pc}{ {\rm P}_{1} }
\nc{\Prb}{ {\rm P} }
\nc{\sd}{\pm \hat{\sigma} }

\nc{\indep}{{\, \perp \! \! \! \perp  \,} }
\nc{\tsps}{^{ {\rm T} } }

\nc{\pu}{\pi_{\rm U}}
\nc{\pbi}{\pi_{\rm B}}
\nc{\pnb}{\pi_{\rm NB}}
\nc{\prp}{\propto}
\nc{\pr}{ {\rm pr} }

\newcommand{\Poi}{\text{Poi}}
\newcommand{\Bin}{\text{Bin}}
\newcommand{\Ber}{\text{Ber}}
\newcommand{\Mult}{\text{Mult}}

\nc{\al}{\alpha}
\nc{\dl}{\delta}
\nc{\la}{\lambda}
\nc{\vep}{\varepsilon}
\nc{\eps}{\epsilon}

\nc{\snf}{\sum_{n=1}^{\infty}}
\nc{\skf}{\sum_{k=1}^{\infty}}
\nc{\sner}{\sum_{n=1}^{86}}
\nc{\sjn}{\sum_{j=1}^{n}}
\nc{\skn}{\sum_{k=1}^{n}}
\nc{\sumim}{\sum_{i=1}^m}
\nc{\sumjn}{\sum_{j=1}^n}
\nc{\sumlL}{\sum_{l=1}^{L}}
\nc{\sumL}{\sum_{l=1}^{L}}
\nc{\sumkK}{\sum_{k=1}^{K_i}}
\nc{\sumrR}{\sum_{r=1}^R}

\nc{\hivp}{\sum_{ {\rm HIV}^+ } }
\nc{\sumiN}{ \sum_{i=1}^N }
\nc{\summM}{ \sum_{m=1}^M }
\nc{\sumjM}{ \sum_{j=1}^M }

\nc{\lsq}{\left[}
\nc{\rsq}{\right]}
\nc{\lbc}{\left \{ }
\nc{\rbc}{\right \} }
\nc{\lp}{\left(}
\nc{\rp}{\right)}

\nc{\imp}{\Rightarrow}
\nc{\lbf}{\lim_{b \rightarrow \infty}}
\nc{\limNinf}{\lim_{N \rightarrow \infty}}
\nc{\limminf}{\lim_{m \rightarrow \infty}}
\nc{\limninf}{\lim_{n \rightarrow \infty}}
\nc{\convd}{\stackrel{D}{\longrightarrow}}
\nc{\convp}{\stackrel{P}{\longrightarrow}}
\nc{\eqd}{\stackrel{{\EuScript D}}{=}}

\nc{\trans}{^{\text T}}
\nc{\ol}{\overline}
\nc{\logit}{\text{logit}\,}

\nc{\rl}{ {\rm {\bf R} } }
\nc{\zah}{ {\rm {\bf Z} } }

\nc{\lkn}{\Lambda^n_k}
\nc{\stp}{ {\cal C}_b }
\nc{\istp}{ {\cal I}_A }
\nc{\snb}{S_{N_b}}
\nc{\stb}{S_{T_b}}
\nc{\ixlog}{I_{ \{ 0 \leq x \leq \log \al \} } }
\nc{\iulog}{I_{ \{ 0 \leq u  \leq \log \al \} } }
\nc{\rgn}{ \Upsilon_n }
\nc{\var}{{\rm var}}
\nc{\cov}{{\rm cov}}
\nc{\corr}{{\rm corr}}
\nc{\dpl}{\partial}
\nc{\half}{ {\textstyle \frac{1}{2}} }
\nc{\tr}{{\rm trace}}
\nc{\real}{\mathbb{R}}
\nc{\bbC}{\mathbb{C}}
\nc{\bbR}{\mathbb{R}}
\nc{\bbone}{\mathbb{1}}
\nc{\bbP}{\mathbb{P}}

\def\boxit#1{\vbox{\hrule\hbox{\vrule\kern6pt\vbox{\kern6pt#1\kern6pt}\kern6pt\vrule}\hrule}}

\nc{\calb}{ {\cal B} }
\nc{\calc}{ {\cal C} }
\nc{\bcalc}{ \mbox{\boldmath{${\cal C}$}}}
\nc{\cald}{ {\cal D} }
\nc{\cale}{ {\cal E} }
\nc{\cali}{ {\cal I} }
\nc{\call}{ {\cal L} }
\nc{\calm}{ {\cal M} }
\nc{\caln}{ {\cal N} }
\nc{\cals}{ {\cal S} }
\nc{\calo}{ {\cal O} }
\nc{\bcalo}{ \mbox{\boldmath{${\cal O}$}}}
\nc{\calt}{ {\cal T} }
\nc{\calv}{ {\cal V} }
\nc{\bcalu}{ \mbox{\boldmath{${\cal U}$}}}
\nc{\calu}{ {\cal U} }
\nc{\calw}{ {\cal W} }
\nc{\calx}{ {\cal X} }

\nc{\sca}{ {\EuScript A} }
\nc{\scb}{ {\EuScript B} }
\nc{\scc}{ {\EuScript C} }
\nc{\scd}{ {\EuScript D} }
\nc{\sce}{ {\EuScript E} }
\nc{\scf}{ {\EuScript F} }
\nc{\scF}{ {\EuScript f} }
\nc{\scg}{ {\EuScript G} }
\nc{\sch}{ {\EuScript H} }
\nc{\sci}{ {\EuScript I} }
\nc{\scj}{ {\EuScript J} }
\nc{\sck}{ {\EuScript K} }
\nc{\scl}{ {\EuScript L} }
\nc{\sclic}{ \scl_i^{\rm c} }
\nc{\scm}{ {\EuScript M} }
\nc{\scn}{ {\EuScript N} }
\nc{\sco}{ {\EuScript O} }
\nc{\scp}{ {\EuScript P} }
\nc{\scq}{ {\EuScript Q} }
\nc{\scr}{ {\EuScript R} }
\nc{\scs}{ {\EuScript S} }
\nc{\sct}{ {\EuScript T} }
\nc{\scu}{ {\EuScript U} }
\nc{\scv}{ {\EuScript V} }
\nc{\scw}{ {\EuScript W} }
\nc{\scx}{ {\EuScript X} }
\nc{\scy}{ {\EuScript Y} }
\nc{\scz}{ {\EuScript Z} }
\nc{\scxo}{ {\EuScript X}_{\rm obs} }
\nc{\Xobs}{ \pmb{\scx}_{\rm obs} }
\nc{\Xcom}{ \pmb{\scx} }
\nc{\Xmis}{ \pmb{\scx}_{\rm mis} }

\nc{\bsci}{ \mbox{\boldmath{$\sci$}}}  
\nc{\bscj}{ \mbox{\boldmath{$\scj$}}}  

\nc{\sumlic}{\sum_{l \in sclic}}

\nc{\scyo}{ {\EuScript Y}_{\rm obs} }

%
%
%

\nc{\bga}{\begin{array}{c}}
\nc{\ena}{\end{array}}

\nc{\mhat}{ {\hat{p}}_M }
\nc{\fhat}{ {\hat{p}}_F }
\nc{\ph} { \hat{p} }

\nc{\ta}{ {\tilde{a}} }
\nc{\tc}{ {\tilde{c}} }

\nc{\bi}{\mbox{\boldmath{$i$}}} 
\nc{\bal}{\mbox{\boldmath{$\alpha$}}} 
\nc{\balpha}{\mbox{\boldmath{$\alpha$}}} 
\nc{\bone}{\mbox{\boldmath{$1$}}} 
\nc{\bbet}{\mbox{\boldmath{$\beta$}}} 
\nc{\bbeta}{\mbox{\boldmath{$\beta$}}} 
\nc{\bDel}{\mbox{\boldmath{$\Delta$}}} 
\nc{\bDelta}{\mbox{\boldmath{$\Delta$}}} 
\nc{\bdel}{\mbox{\boldmath{$\delta$}}}
\nc{\bdelta}{\mbox{\boldmath{$\delta$}}}
\nc{\bet}{\mbox{\boldmath{$\eta$}}}
\nc{\beps}{\mbox{\boldmath{$\epsilon$}}}
\nc{\bvep}{\mbox{\boldmath{$\vep$}}} 
\nc{\bgam}{\mbox{\boldmath{$\gamma$}}} 
\nc{\bgamma}{\mbox{\boldmath{$\gamma$}}} 
\nc{\bGamma}{\mbox{\boldmath{$\Gamma$}}} 
\nc{\boldeta}{\mbox{\boldmath{$\eta$}}} 
\nc{\bLam}{\mbox{\boldmath{$\Lambda$}}} 
\nc{\bLambda}{\mbox{\boldmath{$\Lambda$}}} 
\nc{\blambda}{\mbox{\boldmath{$\lambda$}}} 
\nc{\bmu}{ \mbox{\boldmath{$\mu$}}} 
\nc{\boldnu}{ \mbox{\boldmath{$\nu$}}} 
\nc{\bOm}{ \mbox{\boldmath{$\Omega$}}} 
\nc{\bOmega}{ \mbox{\boldmath{$\Omega$}}} 
\nc{\bom}{ \mbox{\boldmath{$\omega$}}} 
\nc{\bomega}{ \mbox{\boldmath{$\omega$}}} 
\nc{\bpi}{ \mbox{\boldmath{$\pi$}}} 
\nc{\bPi}{ \mbox{\boldmath{$\Pi$}}} 
\nc{\bpsi}{ \mbox{\boldmath{$\psi$}}} 
\nc{\bPsi}{ \mbox{\boldmath{$\Psi$}}} 
\nc{\bphi}{ \mbox{\boldmath{$\phi$}}} 
\nc{\bPhi}{ \mbox{\boldmath{$\Phi$}}} 
\nc{\bxi}{ \mbox{\boldmath{$\xi$}}} 
\nc{\bXi}{ \mbox{\boldmath{$\Xi$}}} 
\nc{\bSig}{\mbox{\boldmath{$\Sigma$}}}
\nc{\bSigma}{\mbox{\boldmath{$\Sigma$}}}
\nc{\bsig}{\mbox{\boldmath{$\sigma$}}}
\nc{\bsigma}{\mbox{\boldmath{$\sigma$}}}
\nc{\btau}{\mbox{\boldmath{$\tau$}}}
\nc{\bThe}{\mbox{\boldmath{$\Theta$}}}
\nc{\bTheta}{\mbox{\boldmath{$\Theta$}}}
\nc{\bthe}{\mbox{\boldmath{$\theta$}}}
\nc{\btheta}{\mbox{\boldmath{$\theta$}}}
\nc{\bzeta}{\mbox{\boldmath{$\zeta$}}}
\nc{\bIm}{\mbox{\boldmath{$\Im$}}}


\nc{\ba}{ { \bf a }}
\nc{\bA}{ { \bf A }}
\nc{\bB}{ { \bf B }}
\nc{\bb}{ { \bf b }}
\nc{\bc}{ { \bf c }}
\nc{\bC}{ { \bf C }}
\nc{\bD}{ { \bf D }}
\nc{\bd}{ { \bf d }}
\nc{\be}{ { \bf e }}
\nc{\bF}{ { \bf F }}
\nc{\boldf}{ { \bf f }}
\nc{\bG}{ { \bf G }}
\nc{\bh}{ { \bf h }}  
\nc{\bH}{ { \bf H }}  
\nc{\bI}{ { \bf I }}  
\nc{\bJ}{ { \bf J }}  
\nc{\bk}{ { \bf k }}  
\nc{\bK}{ { \bf K }}  
\nc{\bL}{ { \bf L }}
\nc{\bM}{ { \bf M }}
\nc{\bn}{ { \bf n }}
\nc{\bO}{ { \bf O }}
\nc{\bP}{ { \bf P }}
\nc{\bp}{ {\bf p }}
\nc{\br}{ { \bf r }}
\nc{\bR}{ { \bf R }}
\nc{\bolds}{ { \bf s }}  
\nc{\bS}{ { \bf S }} 
\nc{\bT}{ { \bf T }} 
\nc{\bt}{ { \bf t }} 
\nc{\bu}{ { \bf u }} 
\nc{\bU}{ { \bf U }}  
\nc{\bv}{ { \bf v }}
\nc{\bV}{ { \bf V }}  
\nc{\bW}{ { \bf W }}  
\nc{\bw}{ { \bf w }}  
\nc{\bx}{ { \bf x }}
\nc{\bX}{ { \bf X }}
\nc{\by}{ { \bf y }} 
\nc{\bY}{ { \bf Y }}  
\nc{\bz}{ { \bf z }} 
\nc{\bZ}{ { \bf Z }}  

\nc{\YR}{[\bY,R]}
\nc{\YgivenR}{[\bY \mid R]}
\nc{\RgivenY}{[R \mid \bY]}
\nc{\Y}{[\bY]}
\nc{\R}{[R]}

\nc{\dio}{d_i^o}
\nc{\timi}{t_{i,m_i}}
\nc{\betahat}{\hat{\bbet}}
\nc{\mui}{\bmu_{\rm I}}
\nc{\mue}{\bmu^{\rm E}}
\nc{\mup}{\bmu^{\rm P}}
\nc{\muihat}{\hat{\bmu}_{\rm I}}
\nc{\muehat}{\hat{\bmu}^{\rm E}}
\nc{\muphat}{\hat{\bmu}^{\rm P}}
\nc{\delhat}{\hat{\bdel}}
\nc{\muhat}{\hat{\bmu}}

\nc{\iid}{\stackrel{\rm iid}{\sim}}
\nc{\law}{\stackrel{\scl}{=}}

\nc{\phiij}{ \phi_{ij}( \Delta_0) }
\nc{\phiiprmj}{ \phi_{i'j}( \Delta_0) }
\nc{\phiijprm}{ \phi_{ij'}( \Delta_0) }
\nc{\phixy}{ \phi( X_i(S_{ik}), Y_j(T_{jl}) ) }
\nc{\phixydo}{ \phi( X_i(S_{ik}), Y_j(T_{jl})-\Delta_0 ) }
\nc{\phixyd}{ \phi( X_i(S_{ik}), Y_j(T_{jl})-\Delta) }
\nc{\phixydstar}{ \phi^*( X_i(S_{ik}), Y_j(T_{jl})-\Delta) }
\nc{\phixystdttil}{ \tilde{\phi}( X_i(s), Y_j(t)-\Delta, \theta) }
\nc{\phixydttil}{ \tilde{\phi}( X_i(S_{ik}), Y_j(T_{jl})-\Delta, \theta) }
\nc{\Nmn}{{\sqrt{N} \over mn}}

\nc{\Xis}{X_i(s)}
\nc{\Yjt}{Y_j(t)}


\nc{\bthehat}{\hat{\bthe}}

\nc{\Ritil}{\tilde{R}_i}

\nc{\Ybar}{\overline{Y}}
\nc{\Rbar}{\overline{R}}
\nc{\Nbar}{\overline{N}}
\nc{\intzeroinf}{\int_0^\infty}

\nc{\Fhat}{\hat{F}}
\nc{\Ghat}{\hat{G}}

\nc{\FhatS}{\hat{F}(S_{ik})}
\nc{\GhatT}{\hat{G}(T_{jl})}

\nc{\Fhatik}{\hat{F}_{ik}}
\nc{\Ghatjl}{\hat{G}_{jl}}
\nc{\Fik}{F_{ik}}
\nc{\Gjl}{G_{jl}}
\nc{\phiijkl}{\phi_{ik,jl}(\Delta)}
\nc{\phiijkltil}{\tilde{\phi}_{ik,jl}(\Delta_0,\theta_0)}
\nc{\ord}{N^{-3/2}}           
\nc{\sumijkl}{\sum_{ijkl}}

\nc{\Citil}{\tilde{C}_i}
\nc{\Crtil}{\tilde{C}_r}
\nc{\Djtil}{\tilde{D}_j}
\nc{\Ditil}{\tilde{D}_i}

\nc{\Cithe}{\tilde{C}^{\theta}_i}
\nc{\Djthe}{\tilde{D}^{\theta}_j}

\nc{\Sikthe}{S_{ik}^{\theta}}
\nc{\Tjlthe}{T_{jl}^{\theta}}

\nc{\Zi}{ \bZ_{-i}}
\nc{\zic}{ \lbc z(\bs_j) \: : \: i \neq j \rbc }
\nc{\zkap}{ \bz_{\kappa} }
\nc{\sumi}{ \sum_i }
\nc{\sumj}{ \sum_j }
\nc{\sumij}{ \sum_{i < j} }
\nc{\sumiandj}{ \sum_{i, j} }
\nc{\zsi}{ z(\bs_i) }
\nc{\Zsi}{ Z(\bs_i) }
\nc{\zsj}{ z(\bs_j) }
\nc{\zsn}{ z(\bs_n) }
\nc{\zsone}{ z(\bs_1) }
\nc{\pZ}{ \Pr \lbc \bZ \rbc }
\nc{\qz}{ Q( \bz ) }
\nc{\qZ}{ Q( \bZ ) }

\nc{\thetaYD}{\theta_{Y\mid D}}
\nc{\thetaD}{\theta_D}
\nc{\psiDY}{\psi_{D\mid Y}}
\nc{\psiY}{\psi_Y}

\nc{\tn}{\Theta^{\nu}}
\nc{\Etn}{E_{\theta^{\nu}}}
\nc{\tnone}{\Theta^{\nu+1}}
\nc{\Lm}{L_{\text{m}}}
\nc{\Lo}{L_{\text{o}}}
\nc{\Ym}{Y_{\text{m}}}
\nc{\Yo}{Y_{\text{o}}}
\nc{\ym}{y_{\text{m}}}
\nc{\yo}{y_{\text{o}}}

\nc{\vijb}{v_{ij} - \bX_{i(j)}  \bbet}
\nc{\vikb}{v_{ik} - \bX_{i(k)}  \bbet}
\nc{\vilb}{v_{il} - \bX_{i(l)}  \bbet}
\nc{\betart}{ \bbet^{(r)}_{t_i} }
\nc{\betarj}{ \bbet^{(r)}_j } 
\nc{\yij}{y_{ij}}
\nc{\Xmisi}{ {\bX_{ i{\rm (mis)} }} }
\nc{\Xobsi}{ {\bX_{ i{\rm (obs)} }} }
\nc{\Zobsi}{ {\bZ_{ i{\rm (obs)} }} }
\nc{\bSigobs}{ \bSig_{  {\rm obs} } }
\nc{\bSigmis}{ \bSig_{  {\rm mis} } }
\nc{\bSigmo}{ \bSig_{  {\rm mis,obs} } }
\nc{\bSigom}{ \bSig_{  {\rm obs,mis} } }
\nc{\Xil}{{\bX}_{il}}
\nc{\Zil}{{\bZ}_{il} }
\nc{\omilr}{\omega_{il}^{(r)}}
\nc{\delio}{\bdel_i^{{\rm obs}} }

\nc{\obs}{{\text{obs}}}
\nc{\mis}{{\text{mis}}}
\nc{\rep}{{\text{rep}}}

\nc{\yio}{ {y_i^{\rm o} }}
\nc{\Yio}{ {Y_i^{\rm o }} }
\nc{\Yim}{ {Y_i^{\rm m} }}
\nc{\yim}{ {y_i^{\rm m} }}
\nc{\Yc}{Y^{\rm c}}
\nc{\Yic}{Y_i^{\rm c}}
\nc{\yc}{y^{\rm c}}
\nc{\yic}{y_i^{\rm c}}

\nc{\yi}{y_i}
\nc{\Yi}{Y_i}

\nc{\fyic}{f ( \yic ; \; \psiY )}
\nc{\fyi}{f ( y_i ; \; \psiY ) }
\nc{\fdigivenyic}{f ( d_i  \mid  \yic ; \; \psiDY )}
\nc{\fditilgivenyic}{f ( \tilde{d}_i  \mid  \yic ; \; \psiDY )}
\nc{\fditilgivenyi}{f ( \tilde{d}_i  \mid  \yi ; \; \psiDY )}
\nc{\Fditilgivenyic}{F ( \tilde{d}_i  \mid  \yic ; \; \psiDY )}
\nc{\Fditilgivenyi}{F ( \tilde{d}_i  \mid  \yi ; \; \psiDY )}
\nc{\fdigivenyi}{f (d_i \mid y_i ; \;  \psiDY  )}
\nc{\fyicdi}{f \left( \yic, d_i \right)}
\nc{\fyidi}{f \left( \yi, d_i \right)}

\nc{\fymidr}{f_{Y \mid R}}
\nc{\fyr}{f_{Y,R}}
\nc{\frmidy}{f_{R \mid Y}}
\nc{\fy}{f_Y}
\nc{\fr}{f_R}

\nc{\fyicgivendi}{f (\yic \mid d_i; \; \thetaYD )}
\nc{\fyigivendi}{f (\yi \mid d_i; \; \thetaYD )}
\nc{\fyicgivens}{f (\yic \mid s; \; \thetaYD )}
\nc{\fyigivens}{f (\yi \mid s; \; \thetaYD )}
\nc{\fdi}{f ( d_i; \; \thetaD )}

\nc{\fyicX}{f ( \yic \mid X_i; \; \psiY )}
\nc{\fyiX}{f ( y_i \mid X_i; \; \psiY ) }
\nc{\fdigivenyicX}{f ( d_i  \mid  \yic, X_i ; \; \psiDY )}
\nc{\fdigivenyiX}{f (d_i \mid y_i, X_i ; \;  \psiDY  )}
\nc{\fyicdiX}{f \left( \yic, d_i \mid X_i \right)}

\nc{\fyicgivendiX}{f (\yic \mid d_i, X_i; \; \thetaYD )}
\nc{\fyigivendiX}{f (y_i \mid d_i, X_i; \; \thetaYD )}
\nc{\fdiX}{f ( d_i \mid X_i; \; \thetaD )}

\nc{\Yistar}{\bY_i^*}

\nc{\Dio}{D_i^{\rm obs}}
\nc{\bdelio}{\bdel_{ i \, {\rm (obs)}} }

\nc{\fygivend}{f_{Y \mid \delta}}
\nc{\fyd}{f_{Y, \delta}}
\nc{\fd}{f_\delta}
\nc{\FD}{F_D}
\nc{\fygivenbd}{f_{Y\mid b, \delta}}

\nc{\alphahat}{\hat{\bal}}
\nc{\phihat}{\hat{\bphi}}
\nc{\thetahat}{\hat{\bthe}}
\nc{\thetatilde}{\tilde{\bthe}}
\nc{\scoretheta}{\bS(\bthe; \, \scc)}
\nc{\hesstheta}{\bH(\bthe; \, \scc)}
\nc{\infotheta}{\sci(\bthe; \, \scc)}
\nc{\sitheta}{\bs_i(\bthe; \, \scc_i)}
\nc{\sithetahat}{\bs_i(\thetahat; \, \scc_i)}

\nc{\loglikobs}{\ell_{{\rm o}}(\bthe; \, \sco)}
\nc{\scoreobs}{\bS_{{\rm o}}(\bthe; \, \sco)}
\nc{\hessobs}{\bH_{{\rm o}}(\bthe; \, \sco)}
\nc{\infoobs}{\scj_{{\rm o}}(\bthe; \, \sco)}

\nc{\Cil}{\scc_{il}}
\nc{\olog}{\lambda^*(\bthe, \Xobs)}
\nc{\LthetaC}{\scl(\bthe; \, \scc)}
\nc{\LthetaCi}{\scl_i(\bthe; \, \scc_i)}
\nc{\LthetaCil}{\scl_i (\bthe; \, \scc_{il}) }
\nc{\lthetaC}{\ell(\bthe; \, \scc)}
\nc{\lthetaCi}{\ell_i(\bthe; \, \scc_i)}
\nc{\lthetaCil}{\ell_i (\bthe; \, \scc_{il}) }
\nc{\Qtheta}{\scq \left( \bthe \, \left| \,  \bthe^{(r)} \right. \right)}
\nc{\thetar}{\bthe^{(r)}}
\nc{\thetas}{\bthe^{(s)}}
\nc{\alphas}{\bal^{(s)}}
\nc{\psis}{\psi^{(s)}}
\nc{\alphasplusone}{\bal^{(s+1)}}
\nc{\psisplusone}{\bpsi^{(s+1)}}
\nc{\alphapsis}{\left( \alphas, \psis \right)}

\nc{\thetarplusone}{\bthe^{(r+1)}}
\nc{\ologi}{\lambda^*_i(\bthe, \Xobs)}
\nc{\llogi}{\lambda_i \left( \bthe, \tilde{\Xcom}_{il} \right) }

\nc{\scxil}{\tilde{\Xcom}_{il}} 
\nc{\siginv}{\bSig_i^{-1}}

\nc{\fofym}{ f \left( \by_i \mid \bbet_m, \bSig \right) }
\nc{\mphim}{ \phi_M \lsq \bSig^{-1/2}(\by_i - \bX_i \bbet_m) \rsq }
\nc{\mphit}{ \phi_M \lsq \bSig^{-1/2}(\by_i - \bX_i \bbet_{t_i}) \rsq }
\nc{\mphij}{ \phi_M \lsq \bSig^{-1/2}(\by_i - \bX_i \bbet_j) \rsq }
\nc{\mphik}{ \phi_M \lsq \bSig^{-1/2}(\by_i - \bX_i \bbet_k) \rsq }
\nc{\expkerm}{ \exp  \lbc -\half \bu_i(\bbet_m)' \bSig^{-1} \bu_i(\bbet_m)
  \rbc } 
\nc{\expkerk}{ \exp  \lbc -\half \bu_i(\bbet_k)' \bSig^{-1} \bu_i(\bbet_k)
  \rbc } 
\nc{\expkerj}{ \exp  \lbc -\half \bu_i(\bbet_j)' \bSig^{-1} \bu_i(\bbet_j)
  \rbc } 
\nc{\normscorem}{\left( \bX_i' \bSig^{-1} \bX_i \bbet_m - \bX_i' \bSig^{-1}
  \by_i \right) } 
\nc{\normscorej}{\left( \bX_i' \bSig^{-1} \bX_i \bbet_j - \bX_i' \bSig^{-1}
  \by_i \right) } 
\nc{\piti}{ \pi \left( t_i, \bal, \bZ_i\bgam \right) }
\nc{\omij}{ \om_{ij} \left( t_i, \bal, \bZ_i\bgam \right) }
\nc{\phibetak}{ \phi_M(\bbet_k) }
\nc{\phibetaj}{ \phi_M(\bbet_j) }
\nc{\dphidbetak}{ \left. \dpl \phibetak \right/ \dpl \bbet_k }
\nc{\dphidbetakf}{ \frac{ \dpl \phibetak }{ \dpl \bbet_k } }
\nc{\uik}{\bu_i \left( \bbet_k  \right)}
\nc{\mset}{ \{ 0, 1, \ldots, M \} }
\nc{\betasigma}{ \left( \lbc \bbet^{(r)}_t \rbc, \bSig^{(r)} \right) }
\nc{\Thetar}{ \bThe^{(r)} }

\nc{\shatkm}{\hat{S}_{\rm KM}}

\nc{\ds}{\displaystyle}

\nc{\beq}{\begin{eqnarray*}}
\nc{\eeq}{\end{eqnarray*}}

\nc{\beqna}{\begin{eqnarray}}
\nc{\eeqna}{\end{eqnarray}}

\nc{\bct}{\begin{center}}
\nc{\ect}{\end{center}}

\nc{\bds}{\begin{description}}
\nc{\eds}{\end{description}}

\nc{\bit}{\begin{itemize}}
\nc{\eit}{\end{itemize}}
 
\nc{\bnu}{\begin{enumerate}}
\nc{\enu}{\end{enumerate}}

\nc{\bgt}{\begin{table}}
\nc{\bgtb}{\begin{center} \begin{tabular}}
\nc{\entb}{\end{tabular} \end{center} }
\nc{\ent}{\end{table}}

\nc{\ts}{\textstyle}


\newtheorem{as}{{Assumption}}
\newtheorem{thm}{{\bf Theorem}}

\makeatletter 
\newcommand*{\rom}[1]{\expandafter\@slowromancap\romannumeral #1@} 
\makeatother 

\section{Introduction}
Consider evaluating the causal effect of an intervention in a context with the following features: 1) the intervention is defined and measured on one type of observational unit, but 2) outcomes of interest are defined and measured on a second, distinct type of unit. Common examples include educational interventions applied to teachers with outcomes of interest defined on students, social interventions applied at neighborhoods with outcomes defined at the level of the resident, or, as will be the focus of the present discussion, interventions applied at sources of air pollution (e.g., power plants) and health outcomes measured among people at specific population locations (e.g., zip codes).  We refer to such a setting as one of {\it bipartite causal inference}, reminiscent of the two types of nodes in a bipartite graph. Such bipartite structures are commonplace in many fields where interest lies in evaluating the causal effects of an intervention.

Consider a setting of bipartite causal inference augmented with the complexity that interconnectedness among the two types of units gives rise to what has been termed in the causal inference literature {\it interference}, where outcomes for a particular unit depend upon treatments assigned to (possibly many) other units.    We term the combination of these two features as the setting of {\it bipartite causal inference with interference}, which has not, to our knowledge, been previously considered.

Most existing work on causal inference with interference is formalized in the familiar setting with one level of observational unit \citep{halloran_study_1991, halloran_causal_1995, sobel_what_2006, hong_evaluating_2006,  rosenbaum_interference_2007, hudgens_toward_2008, tchetgen_causal_2012, zigler_estimating_2012, verbitsky-savitz_causal_2012, luo_inference_2012, aronow_general_2012, bowers_reasoning_2013, aronow_estimating_2017,  liu_large_2014, ogburn_vaccines_2014, perez-heydrich_assessing_2014, liu_inverse_2016}.  The most  well-studied examples are studies of infectious diseases where vaccinating a person will also reduce the infection risk of others who come into contact with that person \citep{halloran_causal_1995, hudgens_toward_2008,  liu_large_2014, ogburn_vaccines_2014, perez-heydrich_assessing_2014} and the analysis of social networks where interventions can affect a unit directly and also through impact on an individuals' peers. Various estimands have been introduced to describe the effect on a particular unit's outcome due to treatments applied to other units, with terminology including {\it indirect effects}, {\it spillover effects}, {\it contamination effects}, and {\it peer effects}, but the common theme is that interference typically arises because unit-to-unit interactions lead outcomes of some to depend on outcomes (and, by extension, treatments) of others.  Methods for estimation and inference in such settings have considered both randomized and observational settings, with emphasis on settings of so-called {\it partial interference} that leverage assumptions of interference within, but not between, distinct and non-overlapping clusters of units  \citep{sobel_what_2006, hong_evaluating_2006, hudgens_toward_2008, tchetgen_causal_2012, liu_large_2014, perez-heydrich_assessing_2014}.  

Similar formalization of interference problems in the bipartite setting presents challenges that have not been previously considered.  One reason is the required technical distinctions relating to the two types of observational unit; defining estimands and corresponding estimators requires maintenance of the distinction between units where interventions occur and those where outcomes are measured.  What's more, settings of bipartite causal inference with interference likely arise due to underlying scientific phenomena that cannot be described by the type of unit-to-unit outcome dependencies common to the study of infectious diseases or social networks.  In the bipartite setting, interference is more likely a consequence of complex exposure dependencies that describe how the impact of a particular treatment propagates across units. Settings of interference due to complex exposure dependencies have been considered in the setting of one observational unit, albeit with much less focus than settings of unit-to-unit outcome dependencies \citep{rosenbaum_interference_2007, verbitsky-savitz_causal_2012, graham_quantifying_2013}.  

The goal of this paper is to formalize the development of potential-outcomes methods relevant to settings of bipartite causal inference with interference.  We define potential outcomes in this setting and introduce {\it interference mappings} describing the network of interconnectedness between units.  From here we formalize alternatives to the commonly-invoked Stable Unit Treatment Value Assumption and propose several causal estimands unique to the bipartite setting.  The discussion of estimands is intentionally general in order to introduce new types of estimands that could potentially be of interest in the bipartite setting.  Ultimately, we invoke several simplifying assumptions, including a bipartite version of partial interference, to focus on a subset of relevant estimands for which corresponding estimators can be derived from existing inverse probability weighted estimators.   Throughout, we highlight similarities and differences with existing estimands and methods for causal inference with interference in settings with one level of observational unit.  

For illustration, we frame the discussion in the context of evaluating interventions designed to reduce pollution-related health burden by limiting harmful emissions from power plants in the U.S..  The features defining the bipartite structure are that interventions are defined and implemented at the level of the power plant, but key questions for regulatory policy pertain to health outcomes (e.g., cardiovascular hospitalizations) measured at population locations across the country. Unlike in most existing literature on causal inference with interference, the interference in the power plant case is not due to dependent outcomes among locations or people (e.g., one person's hospitalization does not affect another person's risk).  Rather, interference in this case is due to the nature of pollution exposure, which derives from complex processes that render an individual location subject to actions at many power plants and many power plants impacting common sets of locations. 

Ultimately, the development in this paper is designed as a framework for addressing problems and data structures that have not been previously considered alongside the formalization of causal inference with interference.  Explicitly targeting the complexities of interference due to air pollution transport presents the first step towards statistical tools for evaluating air quality control policies that have to date relied on deterministic physical-chemical air quality models that are not validated with observed data.

\section{Motivating Setting: Power Plant Regulatory Policies}
Various compounds emitted from power plants undergo complex chemical and physical processes to form harmful air pollution that is transported across space. This phenomenon is known as {\it pollution transport}.  In light of this phenomenon, existing regulatory assessments use deterministic models of pollution transport to simulate regulatory impacts.  From a statistical perspective, the phenomenon of pollution transport manifests as interference between units, since outcomes at one location are dependent on treatments at many pollution sources located ``upwind'' (although note that pollution transport is generally more complex than just the direction of the wind).  Development of new methods for interference can enhance current regulatory assessments by combining rigorous statistical methodology with state-of-the-art knowledge of pollution transport. 

For example, consider a specific intervention that may or may not be implemented at a power plant, namely, the installation of selective catalytic reduction or selective non-catalytic reduction (SnCR) system, a technology known to reduce emissions of nitrous oxides (\NOx), important precursors to the formation of various types of air pollution known to be associated with adverse health outcomes \citep{papadogeorgou_adjusting_2018, bell_ozone_2004}.  We aim to characterize the extent to which installation of such a SnCR system causally impacts hospitalization rates for cardiovascular disease (CVD) among Medicare beneficiaries. This setting fits the description of bipartite causal inference with interference because: 1) SnCR systems are installed (or not) at individual power plants; 2) CVD hospitalizations are measured at zip codes; 3) CVD hospitalizations at a given zip code depend on the constellation of SnCR systems installed at many upwind power plants and; 4) a given power plant may impact the CVD hospitalizations at multiple zip codes.

\section{Potential Outcomes for Bipartite Causal Inference with Interference}
The defining feature of the bipartite structure is the presence of two distinct types of observational units.  First, define the set of {\it interventional units}, $\mathcal{P} = \{p_1, p_2, \ldots, p_P\}$ to be the available observational units upon which interventions either occur or not.  In the motivating example, $\mathcal{P}$ is a set of $P = 473$ power plants located across the U.S..  For each $p_i \in \mathcal{P}$, let  $A_i=1,0$ denote the presence, absence of an intervention at the $i^{th}$ interventional unit, for example, an indicator of whether a power plant installs a SnCR system.   Let $\bA = (A_1, A_2, \ldots, A_P)$ denote a vector of possible treatment assignments to each of the interventional units in $\mathcal{P}$, with $\ba \in \mathcal{A}(P)$ representing one vector of $2^P$ possible treatment allocations.
Denote covariates measured at the level of the interventional units with $\bW_i$, for $i=1,2,\ldots P$.  

Let $\mathcal{M} = \{m_1, m_2, \ldots, m_M\}$, denote a set of $M$ units of a second type, termed {\it outcome units}.  In the motivating example, $\mathcal{M}$ consists of $M=\numzips$ zip codes located across the U.S..  Let $Y_j$ denote a measured outcome at each of the $j=1,2,\ldots M$ outcome units, for example, the number of CVD hospitalizations among Medicare beneficiaries residing at zip code $m_j$.  Similarly, $\bX_j$ could denote covariates measured at the outcome units, for example, zip code level population demographics.  The salient feature of the bipartite structure is that, without further restrictions or assumptions, interventions are not defined on the outcome units (e.g., a zip code cannot be ``treated'' with a SnCR system), yet outcomes of interest are not defined on the interventional units (e.g., a power plant does not have a hospitalization count). 

Defining potential outcomes for the bipartite setting is notationally analogous to settings of one level of observational unit.  Let  $Y_j(\ba)$ denote the potential outcome that would be observed at outcome unit $m_j$ under treatment allocation $\ba$, for example, the number of CVD hospitalizations that would occur at the $j^{th}$ zip code under a specific allocation of SnCR systems on power plants.  In the most general setting, a unique $Y_j(\ba)$ is defined for every possible $\ba \in \mathcal{A}(P)$. The unique feature of these definitions in the bipartite setting is that $Y_j(\ba)$ are defined for $j=1,2,\ldots M$, but $\ba$ is a vector of length $P$.  

\subsection{Common Simplifications to Bipartite Structures}
In many circumstances, the bipartite structure of the data can be simplified by projecting onto the space of one type of observational unit.  Projecting to the space of $\mathcal{M}$ could follow from linking each $m_j \in \mathcal{M}$ to exactly one $p_i \in \mathcal{P}$ by, for example, assuming that each $m_j$ adopts the treatment status of the closest $p_i$.  Such a reduction would extend the definition of the treatment (originally defined on $\mathcal{P}$) to the level of $\mathcal{M}$, and subsequent development could proceed as though $\mathcal{M}$ were the only observational units.  A similar projection to the space of $\mathcal{P}$ could follow by aggregating measures originally defined at the level of $\mathcal{M}$.  For example, one could consider the CVD hospitalizations among all zip codes within a certain distance of each $p_i \in \mathcal{P}$, and proceed as though $\mathcal{P}$ were the only observational units \citep{papadogeorgou_adjusting_2018, papadogeorgou_causal_2017, kim_bayesian_2017}.  Such simplifications might be appropriate in settings for which it is self-evident which single interventional unit corresponds to a given outcome unit, such as in settings where observations are hierarchically clustered (e.g., students within classrooms).  

Other simplifications to the bipartite structure could follow from changing the definition of the intervention (and the subsequent question of interest).  For example, the intervention could be re-defined to pertain to each $m_j$ as some function of the interventions on $\mathcal{P}$. One such possibility in the power plant example would be defining a zip-code level treatment as a function of the treatment statuses of several power plants, such as the proportion of upwind power plants that installed a SnCR system.  This would be similar to a so-called ``exposure mapping,'' \citep{aronow_estimating_2017} which in this case would transform the goal of estimating causal effects of the intervention inherently defined at the level of $\mathcal{P}$ to estimating effects of the new re-defined treatment at the level of $\mathcal{M}$, which may not correspond to any practicable intervention.  

The development herein is designed to formulate causal estimands when no such simplification is appropriate, as in the power plant example where assigning each zip code to one power plant (or vice versa) would be too simplistic in light of the realities of air pollution transport and interest lies in the effects of specific interventions on individual power plants.

\subsection{Extending to the Bipartite Setting: Interference Mappings and Structured SUTVA}\label{sec:InterferenceMappings}
Formalizing potential outcomes and causal estimands in the bipartite setting requires a reformulation of common assumptions about potential outcomes and interference, in particular the stable unit treatment value assumption (SUTVA).  Towards this goal, we cast the bipartite data structure as a network with two different types of nodes, $p_i \in \mathcal{P}$ and $m_j \in \mathcal{M}$, where edges between $p_i$ and $m_j$ denote that interventions applied at $p_i$ have some bearing on outcomes measured at $m_j$.  We use the term {\it interference mapping} to denote such a network structure.  In the power plant setting, this structure is governed by atmospheric and climatological conditions that transport power plant emissions across space as they transform into population pollution exposure. 

For each outcome unit, let the {\it interference set} be the set of interventional units for which the presence or absence of the intervention may affect outcomes \citep{liu_inverse_2016}, a notion that will be made formal with a reformulated statement of SUTVA.  Let  $t_j^\top = (t_{j1}, t_{j2}, \ldots, t_{jP})$, where $t_{ji}= 1 (0)$ if $p_i$ is in the interference set for $m_j$.  Define the {\it interference mapping} as  $T = (t_1, t_2, \ldots, t_M)^\top$, where $T$ is a $M \times P$ matrix denoting the interference sets for all $m_j \in \mathcal{M}$.  This definition of $T$ essentially amounts to what is often considered as an ``adjacency matrix,'' even though the entries of $T$ in this case can encode more complex relationships between the $p_i, m_j$ than spatial adjacency.  For notational simplicity, we will let $p_i \in \intsetj$ denote all $i$ such that $t_{ji}=1$ and use this to refer to all interventional units in the interference set for a given $m_j$.  In the power plant example, the set of $p_i \in \intsetj$ can be thought of as the set of power plants that are ``upwind'' from the $j^{th}$ location, and we will refer to it as such. Similarly, we will let $m_j \in \intseti$ denote all $j$ such that $t_{ji}=1$ and use this to refer to all outcome units that contain $p_i$ in their interference set.  In the power plant example, this can be thought of as all locations that are ``downwind'' from the $i^{th}$ power plant.  

Let $\Aintsub$ denote the subvector of treatment assignments for the interventional units in the interference set for unit $m_j$, that is, the elements $A_i$ corresponding to $p_i \in \intsetj$.  Let $\Aintsubnoti$ be the treatment assignment subvector for interventional units {\it not} in the interference set for $m_j$. We reformulate the usual SUTVA as follows to formalize the meaning of the interference mapping:\\
{\it Assumption:} Structured SUTVA: For a specified interference mapping, $T$:
\begin{itemize}
\item[(i)] $Y_j(\bA) = Y_j(\bA')$ for all $j$ if $\bA = \bA'$,
\item[(ii)] $Y_j(\bA) = Y_j(\bA')$ for all $j$ when $\Aintsub = \bA'_{\{T_j=1\}}$ or equivalently, $Y_j(\Aintsub, \Aintsubnoti) = Y_j(\Aintsub)$ 
\end{itemize}
Part (ii) of structured SUTVA clarifies that potential outcomes for unit $m_j$ need only be considered in terms of the treatment assignment vector of the $p_i \in \intsetj$.  To simplify notation in the subsequent, we will use the subscript $(-i)$ denoting ``not $i$'' to implicitly refer to all interventional units in a given interference set except for $p_i$.  For example, $Y_j(A_i=a, A_{\noti} = \ba_{\noti})$ will refer to the potential outcome at $m_j$ if $p_i$ receives treatment $a$ and the remainder of interventional units in $\intsetj$, denoted with $\pnoti$, receive treatment vector $\ba_{\noti}$.

\begin{figure}[h]
\subfloat[Clusters with SUTVA ]{
\includegraphics[width = .33\linewidth, trim= 0in 0in 0in 0in clip]{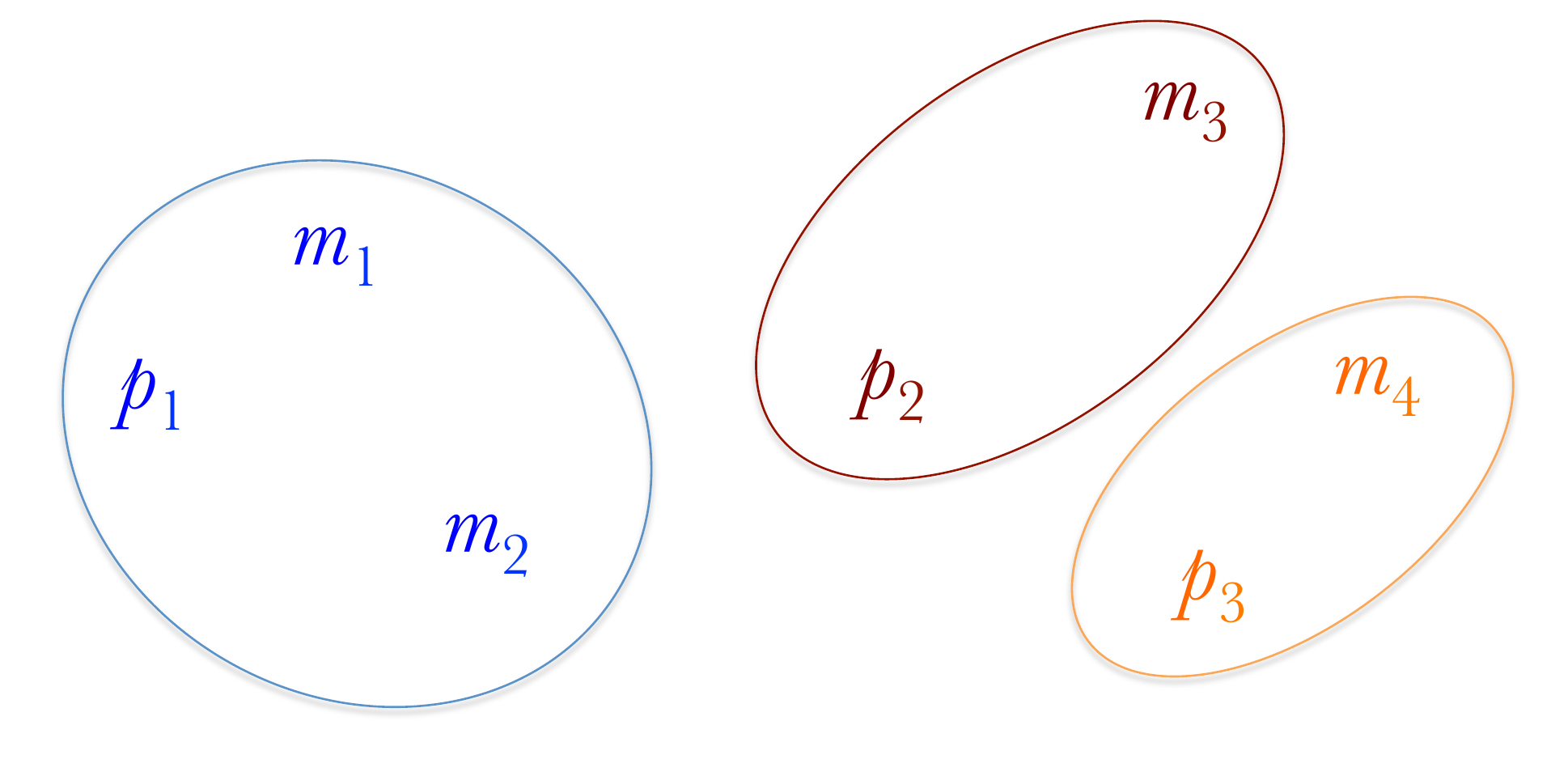}\label{SUTVA}}
\subfloat[Partial Interference]{
\includegraphics[width = .33\linewidth, trim= 0in 0in 0in 0in clip]{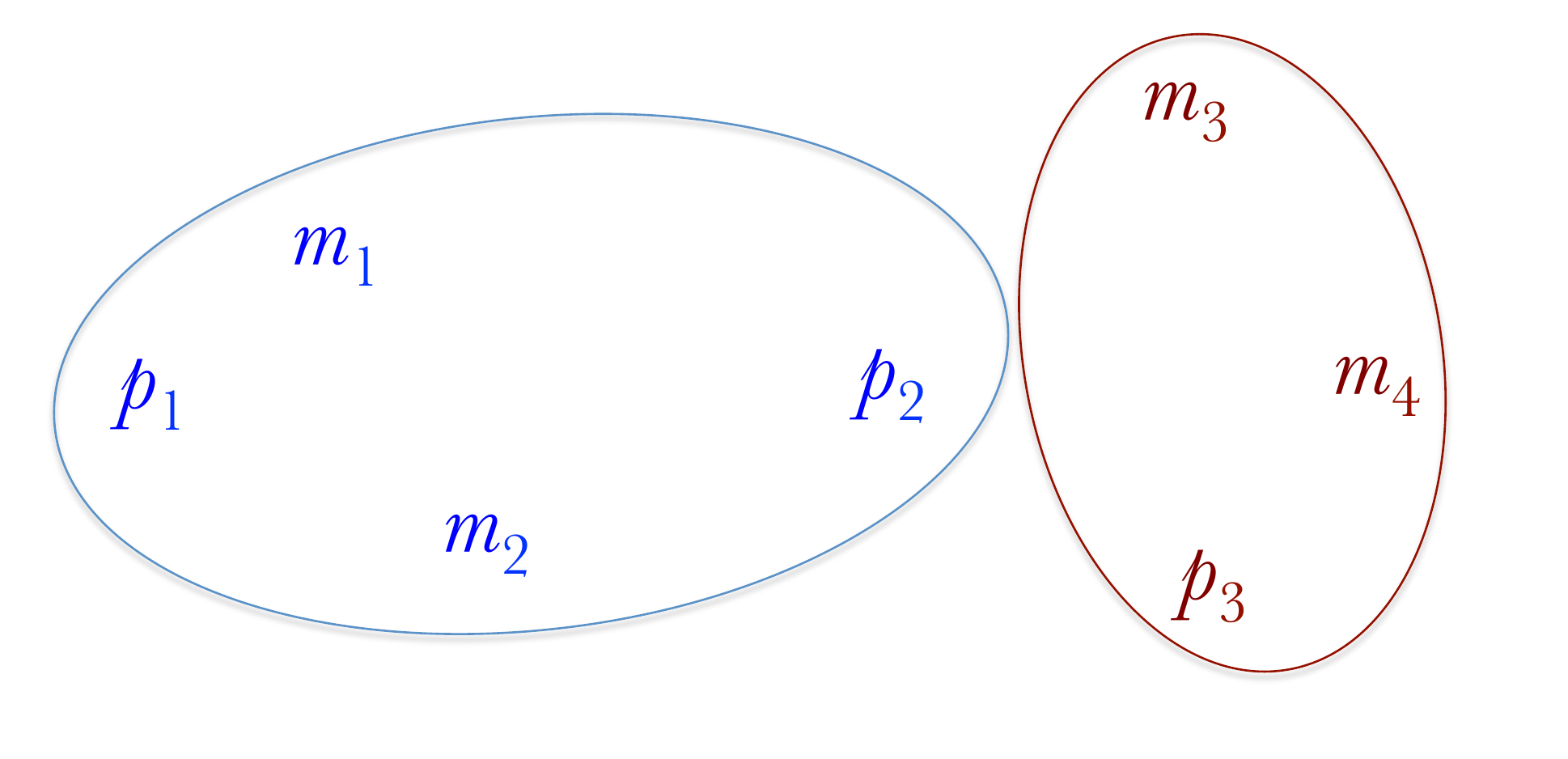}\label{partial}}
\subfloat[General Interference]{
\includegraphics[width = .33\linewidth, trim= 0in 0in 0in 0in clip]{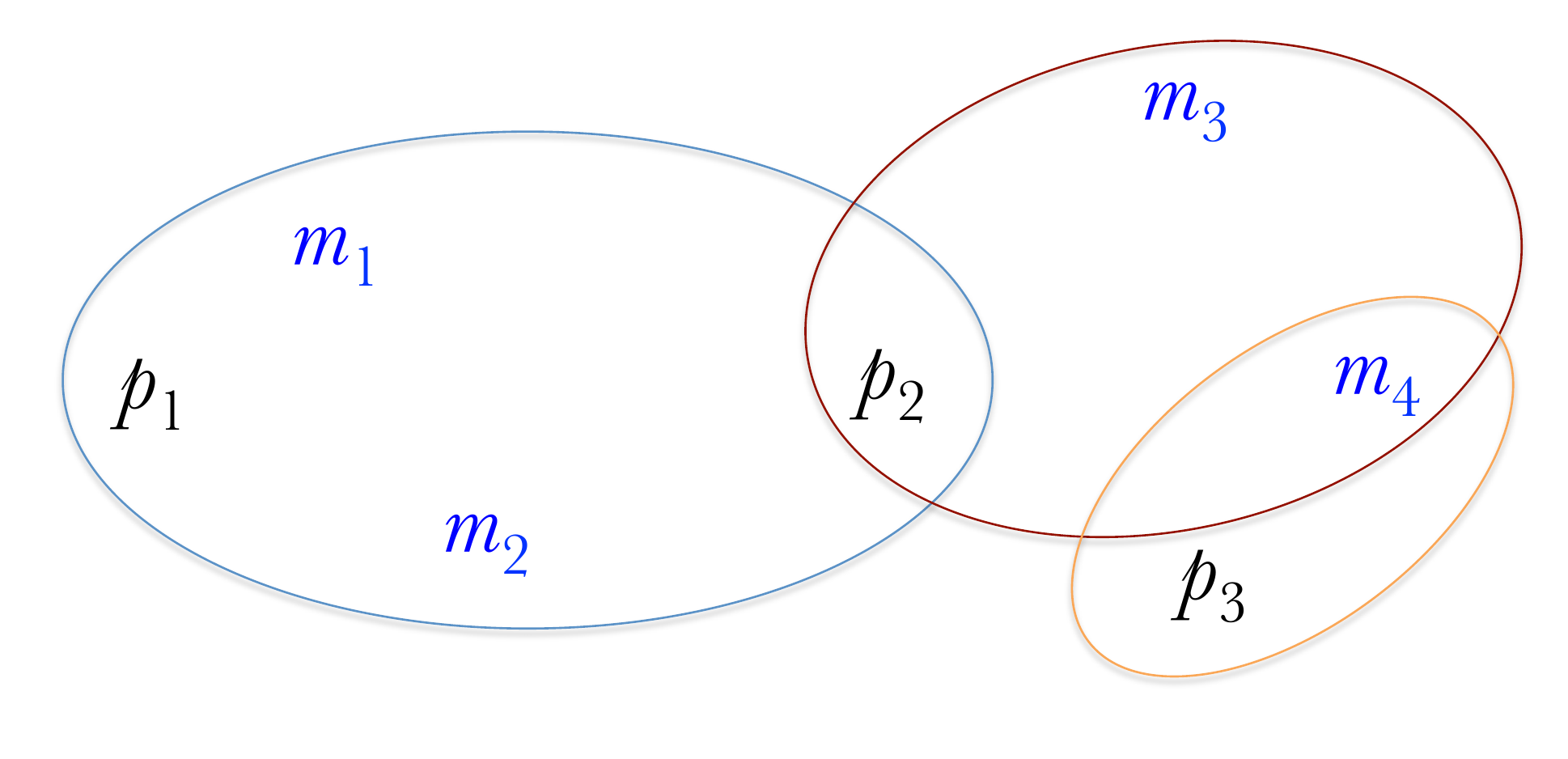}\label{general}}
\caption{Illustrations of interference mappings in simplified setting with $(\mathcal{M} = \{m_1, m_2, m_3, m_4\}$) and ($\mathcal{P} = \{p_1, p_2, p_3\}$).  Potential outcomes at $m_j$ depend upon treatments at all $p_i$ in the same oval. \\[20pt]}\label{mapping}
\end{figure}

Several familiar settings can be formulated via $T$ and Structured SUTVA. To aid illustration, Figure \ref{mapping} schematically depicts three bipartite interference mappings for a simple setting with $M=4$ and $P=3$, where ovals surrounding units represent membership in interference sets.  A setting where outcome units are clustered hierarchically such that each $m_j \in \mathcal{M}$ is subject to exactly one $A_i$ (e.g., students grouped within classrooms) and there is no interference is pictured in Figure \ref{mapping}\subref{SUTVA}. The mapping in this setting is $T  = \left( \begin{smallmatrix} 1 & 0 & 0  \\ 1 & 0 & 0  \\ 0 & 1 & 0  \\ 0 & 0 & 1 \end{smallmatrix} \right)$.  More generally, this type of setting corresponds to $\intsetj$ having exactly one element equal to 1, with every $\intsetj = T_{j'}$ when $m_j$ and $m_{j'}$ are in the same cluster and otherwise $\intsetj^\top T_{j'} = 0$.  The structure depicted in Figure \ref{mapping}\subref{partial} corresponds to a bipartite version of the so-called {\it partial interference} assumption \citep{hudgens_toward_2008, sobel_what_2006}, where: 1) units are divided into non-overlapping clusters consisting of $\ge 1$ outcome unit and $\ge 1$ interventional unit; and 2) outcome-unit potential outcomes are allowed to depend only on the treatments assigned to interventional units within the same cluster.  Figure \ref{mapping}\subref{partial} corresponds to $T  = \left( \begin{smallmatrix} 1 & 1 & 0  \\ 1 & 1 & 0  \\ 0 & 0 & 1  \\ 0 & 0 & 1 \end{smallmatrix} \right)$, and this setting is generally defined by specifying $\intsetj$ as a $P$-vector with $i^{th}$ element equal to 1 for each of the $p_i$ in the same cluster, maintaining the feature that $\intsetj = T_{j'}$ when $m_j$ and $m_{j'}$ are in the same cluster and otherwise $\intsetj^\top T_{j'} = 0$.  Figure \ref{mapping}\subref{general} depicts a more general interference structure that cannot be described by non-overlapping clusters as in the partial interference case, corresponding to $T  = \left( \begin{smallmatrix} 1 & 1 & 0  \\ 1 & 1 & 0  \\ 0 & 1 & 0  \\ 0 & 1 & 1 \end{smallmatrix} \right)$.  Using Figure \ref{mapping}\subref{general} as an example, the set of upwind power plants for unit $m_1$ is $T_1 = (1,1,0)$, and the set of downwind zip codes for unit $p_2$ is $T^\top_2 = (1,1,1,1)$. Note that formulation of interference mappings in the standard single-unit setting could proceed analogously, but with $T$ as $M \times M$ (or $P \times P$),  This would include the most standard setting of no interference, corresponding to $T = diag\{1\}_{M\times M}$.

\section{Estimands for Bipartite Causal Inference with Interference}\label{sec:GeneralEstimands}
As with other settings of causal inference with interference, the interconnectedness between units may not only complicate inference for familiar causal estimands, but may also introduce new causal estimands of interest.  Among causal estimands of frequent interest in the presence of interference with one level of observational unit are so-called ``total'' and ``overall'' effects. We focus in particular on other estimands akin to ``direct effects,'' which capture the effect of an individual unit being treated while holding fixed the treatment statuses of other units in the interference set, and ``indirect'' effects, which capture the effect on a particular unit of holding that unit's treatment status fixed but changing the treatment statuses of others.   We focus on these estimands in particular to explicate complications arising in the bipartite setting due to the fact that treatment is not directly applied or withheld from outcome units, as in studies with one level of observational unit, and notions of ``direct'' and ``indirect'' take on a somewhat different meaning.

Recall that, for a specified interference mapping, $T$, the $(-i)$ subscript denotes all interventional units but $p_i$ within the interference set for a given $m_j$, i.e., $\pnoti$. In principal, causal effects can be defined as comparisons between $Y_j(\ba), Y_j(\ba')$ for any two $\{\ba, \ba'\} \in \mathcal{A}(P)$, that is, any two intervention allocations in the space of possible allocations. As a starting point for development, denote the most primitive individual-level causal effects as:

\begin{equation}
 Y_j(A_i=a, \bA_{\noti}=\ba_{\noti}) - Y_j(A_i=a', \bA_{\noti} = \ba'_{\noti}), \label{ind_ce}
\end{equation}
which denotes the causal effect on outcome unit $m_j$ of treatment allocation $\ba$ with $a_i=a$ {\it versus} treatment allocation $\ba'$ with $a'_i = a'$.  A key feature of the bipartite setting highlighted in (\ref{ind_ce}) is the natural definition of individual effect for every $(p_i,m_j)$ pair for $m_j \in \mathcal{M}$ and $p_i \in \mathcal{P}$.  For example, setting $a=0, a'=1$, and $\ba_{\noti} = \ba'_{\noti}$ in (\ref{ind_ce}) yields a quantity akin to a ``direct'' effect on outcome unit $m_j$ of treating (vs. not treating) interventional unit $p_i$ while holding the treatment status of all other $\pnoti$ fixed at $\ba_{\noti}$. $P$ such ``direct'' effects could be defined for outcome unit $m_j$.

\subsection{Individual-Level Estimands based on Average Potential Outcomes Under Classes of Treatment Allocations}\label{sec:ind_ce}
While development of causal estimands with interference has followed along several lines of development, we adopt a perspective analogous to \cite{hudgens_toward_2008, papadogeorgou_causal_2017}, where estimands are defined based on average individual-level potential outcomes, averaged over many possible treatment allocations.  For example, much work has focused on ``allocation strategies'' representing values of $\ba$ that adhere to a certain probability (or proportion) of treated units, typically denoted with $\alpha$  \cite{tchetgen_causal_2012,liu_large_2014, papadogeorgou_causal_2017, perez-heydrich_assessing_2014}.

We extend this convention and define $\alpha$ to denote a counterfactual treatment allocation strategy where the propensity of interventional units in an interference set to receive treatment $A_i=1$ is set to $\alpha$.  In the bipartite setting, we refer to the definition of $\alpha$ as ``$\mathcal{M}$-centric'' in that it refers to the allocation to all units in the interference set for a particular $m_j$, for example, all power plants ``upwind'' from a specific zip code.  The set of possible treatment allocations adhering to $\alpha$ is denoted with $ \mathcal{A}(\setsizej)$, where $\setsizej$ denotes the number of interventional units in the interference set for $m_j$. 


In the bipartite setting, individual average potential outcomes that average over all treatment allocations fixing $A_i=a$ for a $p_i \in \intsetj$ and having treatment propensity of the interference set fixed to $\alpha$ are defined as:
\begin{equation}
\bar{Y}_j(A_i=a, \alpha) = \sum_{s \in \mathcal{A}(\setsizej -1)} Y_j(A_i=a, \bA_{\noti}=s)\pi(s|A_i=a; \alpha), \label{ind_avg_po}
\end{equation}
where $s \in \mathcal{A}(\setsizej -1)$ \} denotes the set of possible $\ba_{\noti}$ that, along with $a_i = a$, lie in $\mathcal{A}(\setsizej)$.  Here, $\pi(s|A_i=a;\alpha)$ denotes the probability of each such allocation, conditional on $A_i$ being fixed at $a$, which is specified by the researcher to, for example, represent independent Bernoulli allocation of treatments to units or realistic interventions dependent on covariates \citep{tchetgen_causal_2012, papadogeorgou_causal_2017}. Average potential outcomes of the form (\ref{ind_avg_po}) will be used to construct causal estimands of interest.


Using (\ref{ind_avg_po}), we define a bipartite version of an individual-level ``direct effect,'' where ``direct'' is used to refer to the effect of treating (vs. not) a specific $p_i \in \intsetj$, while holding the treatment allocation strategy fixed at $\alpha$: 
\begin{equation}
DE_{(i,j)}(\alpha) = \bar{Y}_j(A_i=1; \alpha) - \bar{Y}_j(A_i=0; \alpha) \label{ind_de}
\end{equation}
For example, $DE_{(i,j)}(\alpha)$ would be the direct effect on outcome unit $m_j$ of treating (vs. not) the $i^{th}$ power plant, when all upwind plants are assigned treatment according to $\alpha$.

Similarly, we define a bipartite version of an individual-level ``indirect effect,'' where ``indirect'' is used to refer to the effect of holding the treatment status of a specific $p_i$ fixed, while changing the allocation to other $\pnoti$:
\begin{equation}
IE_{(i,j)}^a(\alpha, \alpha') = \bar{Y}_j(A_i=a; \alpha) - \bar{Y}_j(A_i=a; \alpha') \label{ind_ie}
\end{equation}
For example, $IE_{(i,j)}^a(\alpha, \alpha')$ would be the indirect effect on outcome unit $m_j$ of holding the treatment status of power plant $p_i$ to $A_i=a$ and changing treatment allocations of other upwind power plants from $\alpha$ to $\alpha'$. 

In addition to expanded notation relative to settings with one level of unit, the salient feature of individual-level effects such as (\ref{ind_de}) and (\ref{ind_ie}) is that they are defined, in full generality, for every $(p_i,m_j)$ pair of $p_i \in \mathcal{P}$ and $m_j \in \mathcal{M}$. This is because, unlike in the single-unit setting, there is no automatic or self-evident notion of which treatment ``directly'' applies to each unit; interest could lie, at least in principle, in the effect of intervening at any power plant on any zip code location. This introduces different strategies for defining the types of average causal effects that will be discussed in Section \ref{sec:Meffects}.  
 
\subsection{$\mathcal{M}$-indexed Average Causal Effects}\label{sec:Meffects}
Recall that, unlike in standard settings of interference where treatments are given directly to one level of unit, the bipartite setting entails no automatic or self-evident notion of which treatment applies directly to which unit.  Thus, it may be of interest to average individual-average potential outcomes for each outcome unit over all interventional units in the interference set.  We introduce the term $\mathcal{M}$-indexed average potential outcomes to refer to average potential outcomes for a given $m_j \in \mathcal{M}$, averaged over $p_i \in \intsetj$: 
\begin{equation}
\bar{Y}_j(a, \alpha) = \frac{1}{\setsizej} \sum_{i \in \intsetj} \bar{Y}_j(A_i=a, \alpha), \label{ind_avg_j}
\end{equation}
which are defined to represent the average potential outcome under $A_i=a$ and allocation program $\alpha$ across all interventional units in the interference set for $m_j$.  

The $\mathcal{M}$-indexed average potential outcomes in (\ref{ind_avg_j}) can be used to define average causal effects paralleling those defined in Section \ref{sec:ind_ce}. Define the $\mathcal{M}$-indexed average direct effect as:
\begin{equation}
DE_j(\alpha) = \bar{Y}_j(1, \alpha) - \bar{Y}_j(0, \alpha) = \frac{1}{\setsizej} \sum_{i \in \intsetj}DE_{(i,j)}(\alpha) \label{de_j}
\end{equation}
to denote the average effect on outcome unit $m_j$ of treating a single $p_i \in \intsetj$ while holding fixed the treatment propensities for all $\pnoti$, averaged over all $p_i \in \intsetj$. The population-average $\mathcal{M}$-indexed direct effect could be defined as $\bar{DE}_\mathcal{M} = \frac{1}{M}\sum DE_j(\alpha)$ representing, for example, the average effect on hospitalizations of installing (vs. not) SnCR on a single upwind power plant while holding the treatment allocation of all upwind plants fixed at $\alpha$.

Similarly, the $\mathcal{M}$-indexed indirect effect is defined as:
\begin{equation}
IE_j^a(\alpha, \alpha') = \bar{Y}_j(a, \alpha) - \bar{Y}_j(a, \alpha') = \frac{1}{\setsizej} \sum_{i \in \intsetj} IE^a_{(i,j)}(\alpha, \alpha') \label{ie_j}
\end{equation}
to represent the average effect on outcome unit $m_j$ of holding treatment at a single $p_i \in \intsetj$ fixed at $a$ while changing the treatment allocation for the interference set from $\alpha$ to $\alpha'$, averaged over all $p_i \in \intsetj$.  The population-average $\mathcal{M}$-indexed indirect effect could be defined as $\bar{IE}_\mathcal{M} = \frac{1}{M}\sum IE_j(\alpha)$ representing, for example, the average effect of holding the SnCR status fixed at an upwind power plant while changing the SnCR allocation of all other upwind plants from $\alpha$ to $\alpha'$.

\subsection{$\mathcal{P}$-indexed Average Causal Effects}\label{sec:Peffects}
The indexing of individual-level potential outcomes in (\ref{ind_avg_po}) (and their corresponding individual-level estimands in (\ref{ind_de}) and (\ref{ind_ie})) by both the outcome unit $j$ and interventional unit $i$ invites averaging potential outcomes over $p_i \in \intsetj$, as in the $\mathcal{M}$-indexed quantities in Section \ref{sec:Meffects}, or averaging potential outcomes over $m_j \in \intseti$, which might be referred to as ``$\mathcal{P}$-indexed'' quantities.  $\mathcal{P}$-indexed effects analogous to (\ref{ind_avg_j}), (\ref{de_j}), and (\ref{ie_j}) could be defined for a particular $p_i$ based on averaging potential outcomes over $m_j \in \intseti$ representing, for example, the average impact of a treatment decision at a particular power plant, averaged across all downwind zip codes.  A main complication with such quantities under the present framework relates to $\alpha$ which, recall, is inherently $\mathcal{M}$-centric in that it refers to the allocation of treatments to interventional units in the interference set for $m_j$, $p_i \in \intsetj$.  Thus, while calculating $\mathcal{M}$-centric average potential outcomes involves fixing $\alpha$ to a single interference set ($\intsetj$),  calculating a $\mathcal{P}$-indexed average potential outcome would correspond to averaging over potential outcomes under multiple different treatment allocations pertaining to the interference sets of each $m_j \in \intseti$, i.e., to fixing the treatment allocation of interventions to all $p_i \in \intsetj$ for all $m_j \in \intseti$.  For example, one could define a $\mathcal{P}$-indexed direct effect analogous to (\ref{de_j}) to characterize how installing an SnCR system at power plant $p_i$ affects hospitalization outcomes, on average across all downwind zip codes, with each downwind zip code having propensity of SnCR installation among its respective upwind plants fixed to $\alpha$.  Such $\mathcal{P}$-indexed effects, while potentially of interest and an important topic for future work, are not pursued here in favor of exploration of a subset of $\mathcal{M}$-indexed effects for which estimators can be derived from existing work.  

\subsection{Key-Associated Average $\mathcal{M}$-indexed Causal Effects}\label{sec:subMeffects}
The fundamental feature that individual-level causal effects can be naturally defined for every $(p_i,m_j)$ pair in the bipartite setting may be simplified in settings where each outcome unit can be associated with a single $p_i \in \intsetj$ at which intervening is of particular interest.  Denote such an interventional unit with $\pstar$, defined for every $m_j \in \mathcal{P}$. We will refer to $\pstar$ as the ``key associated'' interventional unit for outcome unit $m_j$.  In practice, criteria for determining the relevant $\pstar$ for every $m_j \in \mathcal{M}$ will undoubtedly vary, but examples in the power plant setting include the closest or largest power plant located upwind from a given location. When indexing other quantities defined for $\pstar$, we will simplify notation and use the subscript $i^*$.  For example, $A_{i^*}$ will be used to denote the treatment assignment of $\pstar$.

The potential outcomes and estimands in Section \ref{sec:Meffects} averaged over all interventional units for each $m_j$, owing to the fact that the bipartite setting does not inherently contain a notion of which $p_i$ corresponds ``directly'' to each $m_j$.  However, definition of a $\pstar$ for every $m_j \in \mathcal{M}$, invites focus on only a subset of the individual-level causal effects of type (\ref{ind_de}) and (\ref{ind_ie}), specifically those corresponding to the intrinsic interest in the key-associated interventional unit.  Rather than consider every $(p_i,m_j)$ pair, interest is confined to exactly one individual-level direct effect ($DE_{(i^*,j)}(\alpha)$) and exactly one individual-level indirect effect ($IE^a_{(i^*,j)}(\alpha, \alpha')$) for each $m_j \in \mathcal{M}$.

Population-average analogs of these effects can be defined as: 
\begin{eqnarray}
\bar{DE}^*(\alpha) = \frac{1}{M} \sum_{j=1}^M DE_{(i^*, j)}(\alpha) \label{de*} \\
\bar{IE}^{*a}(\alpha, \alpha') = \frac{1}{M} \sum_{j=1}^M IE^a_{(i^*, j)}(\alpha, \alpha') \label{ie*}.
\end{eqnarray}
The estimand (\ref{de*}) corresponds to the average effect on outcome units of treating (vs. not) the key-associated unit while holding fixed the allocation program to other interventional units in the interference set.  In the power plant example, this could correspond, for example, to the average effect on hospitalizations of installing an SnCR system on the closest power plant while holding fixed the allocation of SnCR systems to other upwind plants.  The estimand (\ref{ie*}) corresponds to the average effect on outcome units of holding the treatment at the key-associated unit fixed while varying the allocation program to other interventional units in the interference set from $\alpha$ to $\alpha'$.  In the power plant example, this could correspond to the average effect on hospitalizations of holding the SnCR status of the closest power plant fixed while changing the allocation to other upwind plants.

\section{Estimators under Bipartite Partial Interference in Observational Studies}
\begin{figure}[t]
\centering
\subfloat[]{
\includegraphics[width=0.52\textwidth]{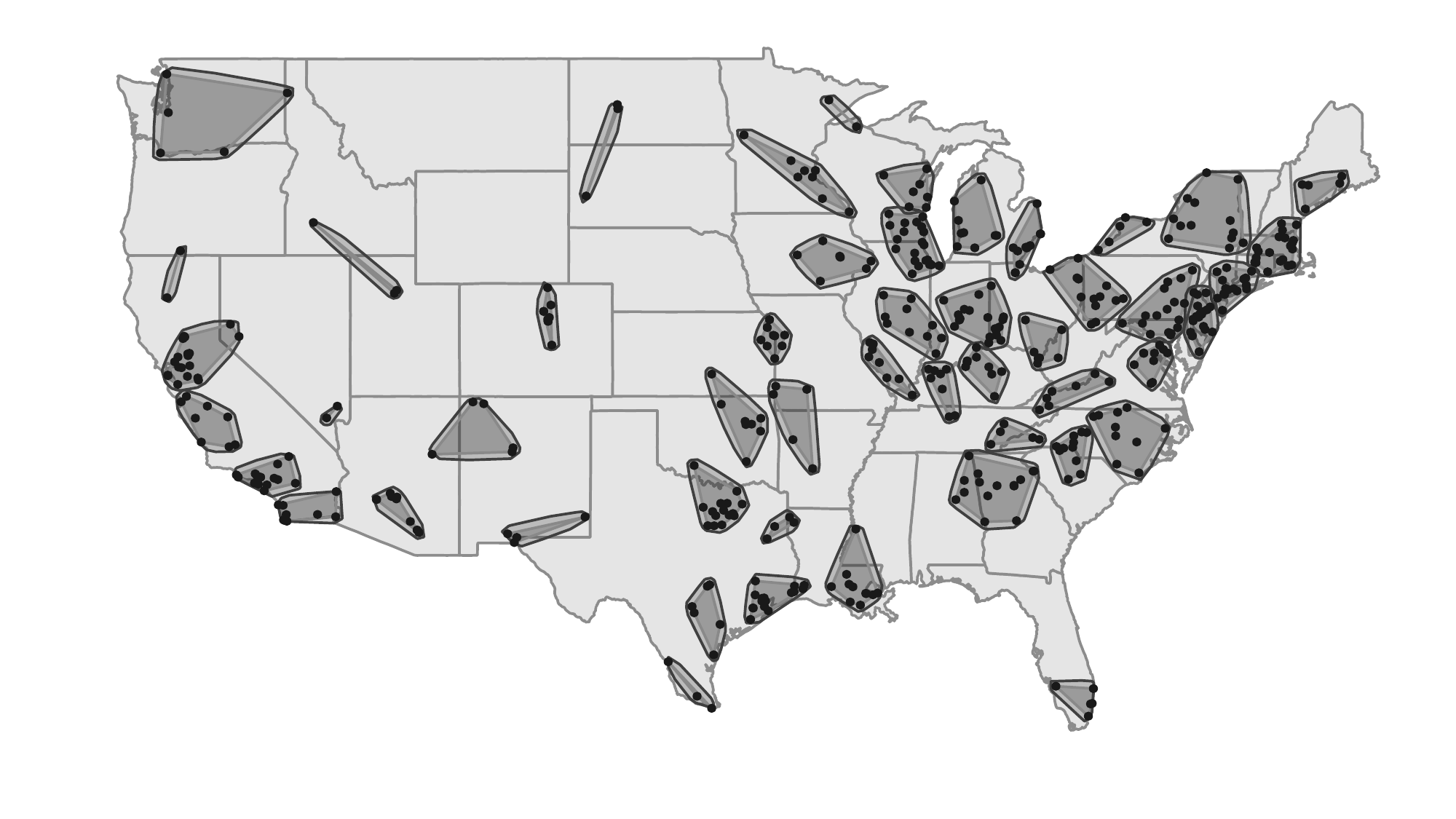}
\label{fig:app_all_clusters}}
\subfloat[]{
\includegraphics[width=0.38\textwidth]{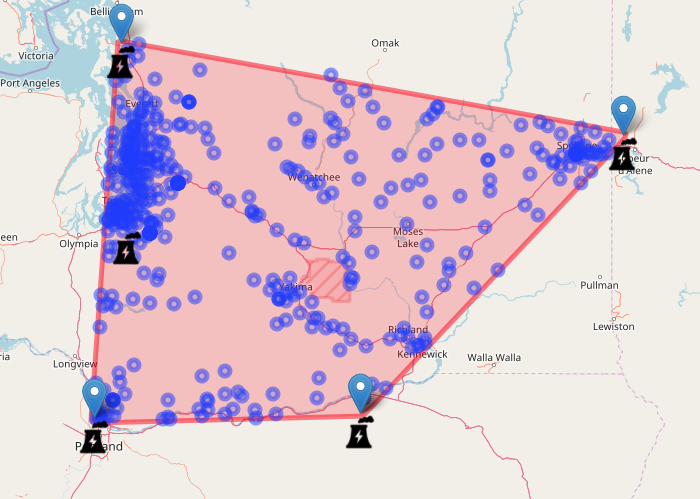}\label{fig:app_one_cluster}}
\caption{(a) Grouping of power plants in interference clusters and assignment of zip codes to clusters. Each cluster is depicted with two polygons, the inner polygon corresponds to the convex hull of the power plants, and the outer polygon corresponds to the convex hull of zip code centroids in that cluster. (b) One cluster of power plants and corresponding zip codes with zip codes' centroids depicted in blue.}\label{fig:app_clusters}
\end{figure}

While the development in Section \ref{sec:GeneralEstimands} pertains to a general form of interference mappings, $T$, we illustrate the development of bipartite estimators for the simplified setting of {\it partial interference}, for which existing estimators in the one unit setting extend in a relatively straightforward way.  Consider a partition of $\mathcal{P}$ into $K$ non-overlapping clusters of interventional units: $\{P^1, P^2, \ldots, P^K\}$, each of size $\pclustsize$.  For example, power plants could be clustered according to geographic proximity.   Consider a corresponding grouping of $\mathcal{M}$ into exactly $K$ non-overlapping clusters $\{M^1, M^2, \ldots, M^K\}$, where each $M^k$ consists of $\mclustsize$ outcome units that are linked in some fashion to the interventional units in $P^k$. For example, $M^k$ could consist of all of zip code locations within a certain distance of at least one of the power plants in $P^k$. Partial interference in this case assumes that potential outcomes at $m_j \in M^k$ depend only on the treatments assigned to  $p_i \in P^k$.  In the terminology of Section \ref{sec:InterferenceMappings}, this amounts to an interference mapping where $T$ has a block structure such that, for $k=1,2,\ldots,K$, $\intsetj$ is the same for all $m_j \in M^k$, with $t_{ji}=1$ for all $p_i \in P^k$ and $t_{ji}=0$ otherwise. Recall that this implies $\intsetj^\top  T_l = 0$ for all $m_j \in M^k, m_l \in M^{k'}$, denoting no common interventional units in the interference sets for two outcome units in different clusters. For simplicity, assume that for each $k=1,2,\ldots,K$, both $M^k$ and $P^k$ contain at least one unit of their respective type. Figure \ref{fig:app_clusters} illustrates one such clustering in the power plant example.

\subsection{Cluster-Level Average Potential Outcomes under Partial Interference}
The partial interference assumption invites definition of cluster-specific analogs to the average effects proposed in Section \ref{sec:Meffects}.  The expressions $p_i \in \intsetj$ and $m_j \in \intseti$ become equivalent to $p_i \in P^k$ and $m_j \in M^k$, and $\alpha$ takes on the familiar meaning of the cluster-level treatment propensity referring to all $p_i \in P^k$. $\mathcal{M}$-indexed effects such as those in (\ref{de_j}) and (\ref{ie_j}) could be averaged over all $j \in M^k$ for all $k = 1, 2, \ldots, K$ to create cluster averages.  However, we focus on developing estimators for the power plant setting that correspond to analogs to the key-associated $\mathcal{M}$-indexed estimands defined in Section \ref{sec:subMeffects}.

Specifically, based on (\ref{ind_avg_po}) we define cluster-level average potential outcomes of the form:
\begin{equation}
\bar{Y}^k(A_{i^*} = a, \alpha) = \frac{1}{\mclustsize} \sum_{j \in M^k} \bar{Y}_j(A_{i^*}=a, \alpha) \label{cluster_avg_po}
\end{equation}
to denote the cluster-level average potential outcome when $\pstar$ receives treatment $a$ and all other $p_i$ in the cluster receive allocation program $\alpha$, averaged over all outcome units in the cluster. Population average potential outcomes can be subsequently defined with $\bar{Y}(A_{i^*}=a, \alpha) = \sum_{k} \bar{Y}^k(A_{i^*}=a, \alpha) / K$.   

Formulation of cluster-average potential outcomes leads to the following expressions for cluster-level average direct and indirect effects:
\begin{eqnarray}
DE^{k*}(\alpha) = \bar{Y}^k(A_{i^*} = 1, \alpha) - \bar{Y}^k(A_{i^*} = 0, \alpha) = \frac{1}{\mclustsize}\sum_{j \in M^k} DE_{(i^*,j)}(\alpha) \label{de_k*} \\
IE^{k*a}(\alpha, \alpha') = \bar{Y}^k(A_{i^*} = a, \alpha) - \bar{Y}^k(A_{i^*} = a, \alpha') = \frac{1}{\mclustsize}\sum_{j \in M^k} IE^a_{(i^*,j)}(\alpha) \label{ie_k*} 
\end{eqnarray} 
Population-level effects defined in (\ref{de*}) and (\ref{ie*}) can be constructed from (\ref{de_k*}) and (\ref{ie_k*}) as
\begin{align}
DE^*(\alpha) = & \frac{1}{K} \sum_kDE^{k*}(\alpha), \text{ and } \label{de^*} \\
IE^{a*}(\alpha, \alpha') = & \frac{1}{K} \sum_k IE^{ka*}(\alpha). \label{ie^*}
\end{align}

\subsection{IPTW Estimator for Average Potential Outcomes}\label{sec:ipwestimator}
Here we illustrate that, among all the estimands defined for the bipartie setting in Section \ref{sec:GeneralEstimands}, existing estimators in \cite{tchetgen_causal_2012, liu_inverse_2016, papadogeorgou_causal_2017} are essentially directly applicable to estimands that rely on: 1) clusters of units and partial interference and 2) a relevant key-associated $\pstar$ defined for each $m_j \in \mathcal{M}$.  Technical development follows from previous work, ensuring that population (cluster) quantities related to treatment assignment are confined to $i=1,2,\ldots,P (P^k)$ while population (cluster) quantities related to outcomes are confined to $j=1,2, \ldots, M (M^k)$. Otherwise, theoretical underpinnings of the estimators extend trivially.  

Specifically, we propose a refinement (to reflect the bipartite setting) of the simple estimator proposed in \cite{tchetgen_causal_2012} for the cluster-level average potential outcomes in (\ref{cluster_avg_po}). A corresponding estimator for the population-level average potential outcome follows immediately, with asymptotically normal distribution as the number of clusters $K$ increases to infinity.  This development follows existing work in \cite{papadogeorgou_causal_2017, liu_inverse_2016, tchetgen_causal_2012}, leading directly to estimators for the population-level key-associated $\mathcal{M}$-indexed direct and indirect effects in (\ref{de^*}) and (\ref{ie^*}) with known asymptotic distributions.  

The estimator for the cluster-level average potential outcome has the familiar form:
\begin{equation}
\widehat Y^k(A_{i^*}=a; \alpha) = \frac1{|M^k|} \sum_{j \in M^k}\frac{\numerator}{\denominator}  I(A_{i^*} = a) Y_j,
\label{eq:group_estimator}
\end{equation}
with corresponding estimator for the population-average potential outcome:
\begin{equation}
\widehat Y(A_{i^*} = a;\alpha) = \frac1K \sum_{k = 1}^K \widehat Y^k(A_{i^*}=a; \alpha)
\label{eq:pop_estimator}
\end{equation}
The term $\denominator$ in the denominator of (\ref{eq:group_estimator}) represents the cluster-level propensity score for the probability that the $p_i \in P^k$ receive the observed treatment vector $\bA^k$, conditional on the interventional-unit and outcome unit covariates in the cluster, $\bW^k$ and $\bX^k$.   The term $\numerator$ in the numerator of (\ref{eq:group_estimator}) represents the user-specified probability distribution of different cluster-level treatment allocations adhering to the program $\alpha$ (specified in accordance with (\ref{ind_avg_po})).  

Under the following assumptions and following work in \cite{tchetgen_causal_2012, liu_inverse_2016, papadogeorgou_causal_2017}: $\widehat Y^k(A_{i^*}=a; \alpha)$ in (\ref{eq:group_estimator}) is unbiased for $\bar{Y}^k(A_{i^*} = a, \alpha)$ in (\ref{cluster_avg_po}) when the cluster propensity score is known; unbiasedness of $\widehat Y(A_{i^*}=a;\alpha)$ in (\ref{eq:pop_estimator}) for $\bar{Y}(A_{i^*}=a, \alpha)$ follows trivially.

\begin{assumption}
\textit{Positivity.} For $k \in \{1, 2, \dots, K\}$, the probability of observing cluster treatment vector $\bA^k = \ba^k$ given cluster covariates $\bW^k, \bX^k$ is denoted by $f_{\bA| \bW, \bX, k}(\bA^k = \ba^k | \bW^k, \bX^k)$ and is positive for all $\ba^k \in \mathcal{A}(\pclustsize)$. 
	\label{ass:group_positivity}
\end{assumption}

\begin{assumption}
\textit{Ignorabililty} For $k \in \{1, 2, \dots, K \}$, the observed cluster treatment $\bA^k$ is conditionally independent of the set of cluster potential outcomes $\bY^k(\cdot)$ given the cluster covariates $\bW^k, \bX^k$, denoted as $\bA^k \amalg \bY^k(\cdot) | \bW^k, \bX^k$.
\label{ass:ignorability}
\end{assumption}

Under superpopulation (of clusters) versions of Assumption \ref{ass:group_positivity} and Assumption \ref{ass:ignorability} as stated in \cite{papadogeorgou_causal_2017}, $\widehat Y(A_{i^*}=a, \alpha)$ is consistent and asymptotically normal for the superpopulation counterpart to the above estimands for a known or correctly specified and estimated parametric propensity score model ($\denominator$)



\section{Evaluating SnCR Systems on Medicare Hospitalizations in the Presence of Pollution Transport}

A previous analysis that simplified the bipartite structure by projecting to the level of $\mathcal{P}$ showed that SnCR systems at coal- or gas-fired power plants reduce ambient air pollution in the areas immediately surrounding power plants and in other ``downwind'' areas \citep{papadogeorgou_causal_2017}.  Here we conduct an analysis of SnCR on  hospitalizations with a more complete regard for the bipartite structure of the problem.  Specifically, we estimate direct and indirect effects (\ref{de^*}) and (\ref{ie^*}) of SnCR installation on zip code hospitalizations for CVD among Medicare beneficiaries.

The set of interventional units consists of 473 coal or natural gas burning power plants operating in the continental U.S. during the summer months (June-August) of 2004. These power plants, with their characteristics and important aggregate area-level characteristics (i.e., $\bW$) have been previously described in detail \citep{papadogeorgou_adjusting_2018}. Power plants are partitioned into 50 clusters as in \cite{papadogeorgou_causal_2017}.

\begin{figure}[!t]
\centering
\includegraphics[width = 0.6\textwidth]{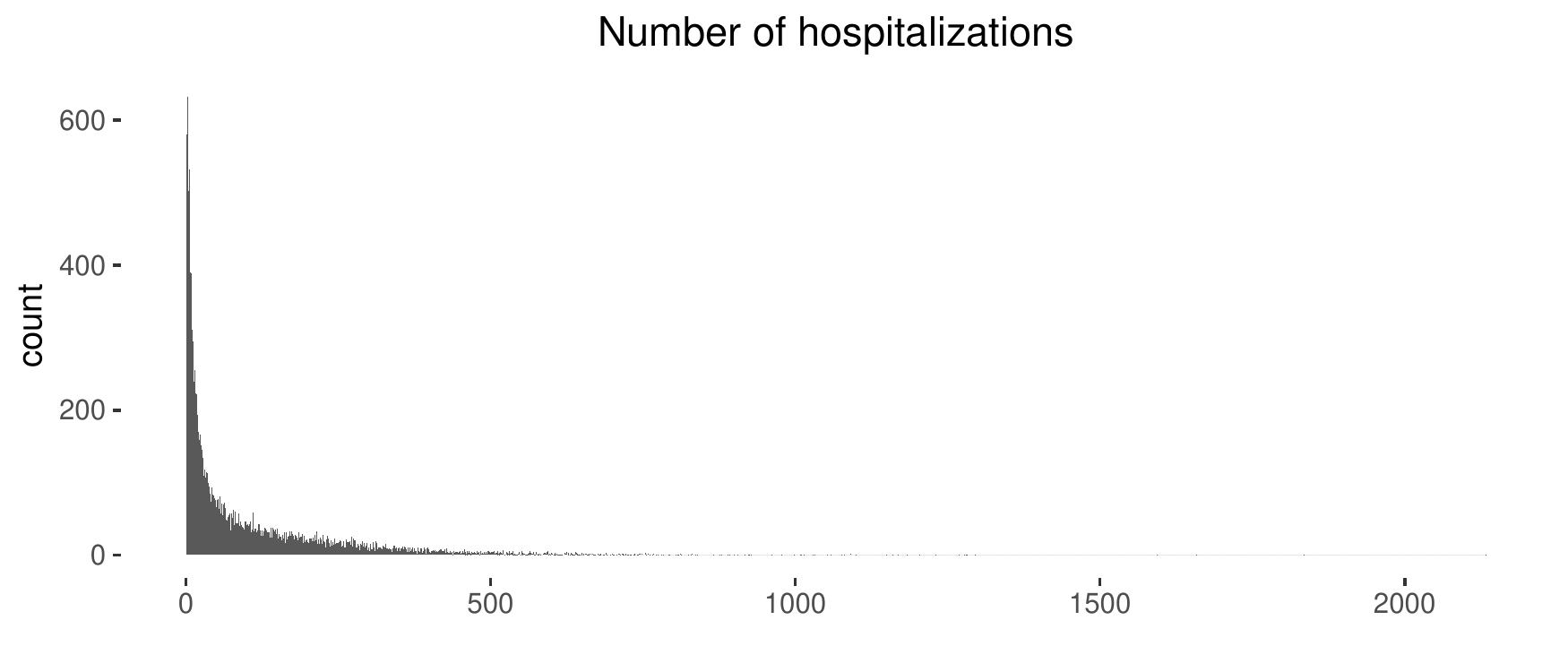}
\caption{Distribution of observed 2005 number of cardiovascular hospitalizations for the \numzips zip codes included in the analysis.}
\label{fig:cardio_rate}
\includegraphics[width = 0.85 \textwidth]{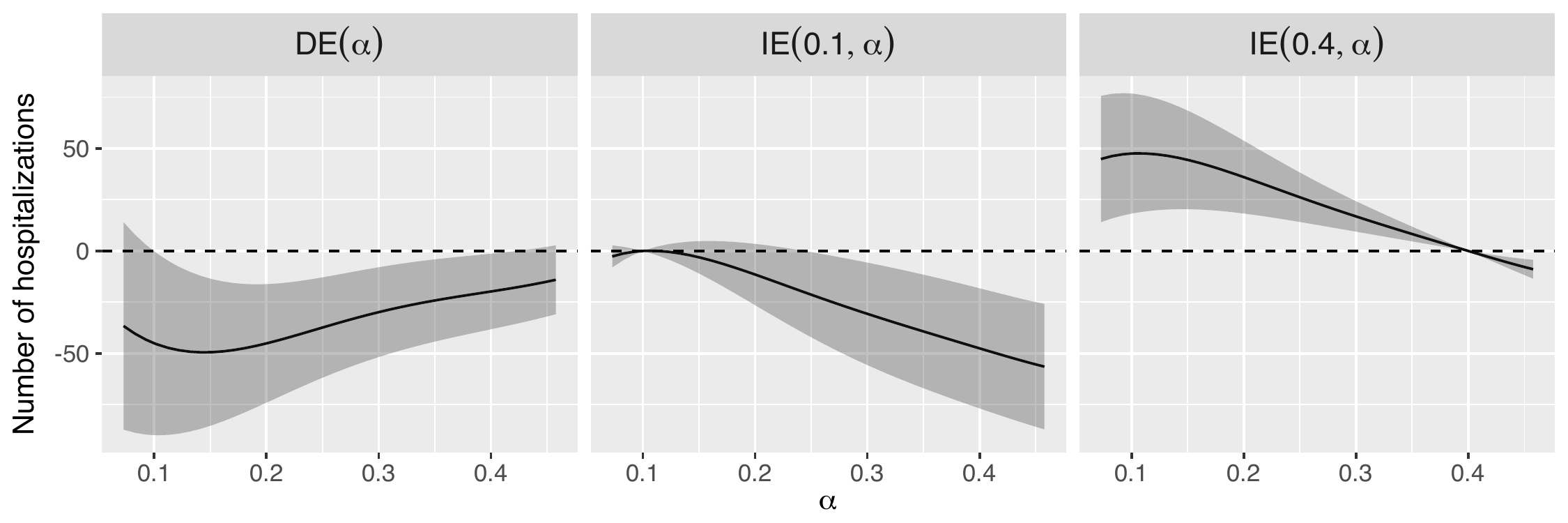}
\caption{Direct effect and indirect effect estimates and confidence intervals.}
\label{fig:app_results}
\end{figure}

The initial set of outcome units considered for this analysis corresponds to 37,240 U.S. zip codes, each with a measured number of hospitalizations for cardiovascular disease (codes ICD-9 390 to 459) among Medicare fee-for-service beneficiaries in 2005 (no outcome-unit covariates, $\bX$, are included in the analysis). Zip codes were assigned to a cluster of power plants if the zip code centroid was located within the area defined by the power plant locations' convex hull and a buffer zone of 30km. If a zip code belonged to more than one cluster based on this definition, it was assigned to the cluster that included the closest power plant.  If a zip code was not within 30km of the buffer zone of any power plant cluster, it was excluded from the analysis. This resulted in a total of \numzips  zip codes representing the population of interest of areas of the U.S. that are considered likely to be impacted by interventions at power plants (See Figure \ref{fig:app_clusters}).  Figure \ref{fig:cardio_rate} shows the observed distribution of the hospitalization outcome over the \numzips zip codes.  

For each zip code, $m_j$, the key-associated power plant, $\pstar$, is defined to be the power plant located closest to the centroid of the zip code. Corresponding key-indexed direct and indirect effects thus cohere to notions of intervening to control local pollution (e.g., from the closest plant) vs. those to control long-range transported pollution from more distant upwind plants, which are important distinctions for development of local and interstate (or regional) regulatory policies.  Key-indexed direct and indirect effects were estimated using the IPW estimator defined in Section \ref{sec:ipwestimator} for values of $\alpha$ ranging from the 20$^{th}$ to the 80$^{th}$ percentile of the observed proportion of treated power plants across the clusters.  The propensity score model was specified as a logistic regression adjusting linearly for power plant, weather, and demographic covariates and including a cluster-specific random effect to match the previous analysis of \cite{papadogeorgou_causal_2017}.  Power plant characteristics include the percent of total capacity at which a plant typically operates, the amount of fuel energy burned, an indicator of Phase 2 participation in the Acid Rain Program, an indicator for whether a plant burns mostly gas fuel, and indicators for the size of the plant in terms of number of generating units.  Area-level characteristics include ambient temperature, median household income, median house value, population per square mile, and population percentages of high school graduates, residence in urban areas, white, black, and hispanic populations, housing occupancy, poverty, and migration to the area within 5 years. 

The numerator specifying counterfactual treatment allocation probabilities was specified as independent Bernoulli assignments to treatment
\( \displaystyle \numerator =
\prod_{p_{k\neq i} \in \intsetj} \alpha^{A_k}(1 - \alpha)^{1 - A_k} \)
as in \cite{tchetgen_causal_2012}.
Results are depicted in Figure \ref{fig:app_results}. The direct effect is estimated to be negative for all values of $\alpha$ (achieving statistical significance at the 0.05 level for all $\alpha \geq 0.1 $), implying that installation of SnCR at a zip code's closest power plant leads to a significant reduction in number of cardiovascular hospitalizations at that location. Note that the direct effect becomes less pronounced as  $\alpha$ increases, indicating that installing SnCR on a zip code's closest power plant has a smaller impact on CVD hospitalizations when more upwind power plants also have SnCR installed. The other two plots in Figure \ref{fig:app_results} depict estimates of the indirect effect $IE^{0*}(\alpha_1, \alpha_2)$ for values $\alpha_1 \in \{0.1, 0.4\}$. These results represent expected changes in hospitalizations when the closest power plant does not have SnCR and the propensity of upwind power plants to install SnCR shifts from $\alpha_1$ to $\alpha_2$. The decreasing trend in both plots indicates that a higher proportion of SnCR among upwind plants leads to decreased CVD hospitalizations when the closest power plant remains without SnCR. For example, a change in the propensity of upwind units to install SnCR from 10\% to 45.8\% would lead to 56.4 (95\% CI: $25.8 - 87.1$) fewer hospitalizations on average per zip code when the closest plant remains without SnCR.

Overall, the results of the analysis indicate the benefit of installing SnCR for reducing CVD hospitalizations among Medicare beneficiaries with a careful account of how the effectiveness of controls installed at nearby power plants interacts with interventions at upwind plants. 
 
\section{Discussion}
We have introduced the new setting of bipartite causal inference with interference, which arises in a variety of settings where interventions are enacted at one type of observational unit, outcomes of interest are defined and measured at a distinct type of unit, and the complexities of exposure patterns lead outcomes to depend on the treatments of many interventional units.  The setting is particularly relevant to the study of air pollution regulatory policy, where interventions occur at pollution sources (e.g., power plants), health outcomes are measured at population locations (e.g., zip codes), and complex atmospheric processes and long-range pollution transport lead to interference.  Formalization of this setting represents an important added dimension to recent work on interference that extends the formality of potential outcomes methods to settings that do not cohere to the oft-considered setting of one level of observational unit and unit-to-unit outcome dependencies (e.g., infectious diseases or social networks).

Potential outcomes and causal estimands were formulated generally, drawing commonalities and distinctions with existing work for interference.  While the general development of estimands was designed to introduce the possibilities of formalizing the bipartite interference problem, estimators were developed for only a subset of possible estimands.  For illustration and to motivate the use of the bipartite framework in an applied problem, we ultimately employed estimators that rely on the assumption of partial interference and require that interest lies in a single key-associated interventional unit for each outcome unit.  The proposed estimators relied heavily on existing work developed in the one-unit setting.  Future research to expand beyond these simplified estimators is clearly warranted, including formulations that go beyond the perspective of individual-average potential outcomes averaged over specified allocation programs (e.g., as in \cite{forastiere_identification_2016}).

Even with the simplifications that led to the proposed estimators, the formalization of bipartite interference and application of the simplified estimators in the context of the power plant evaluation represents an important step in air pollution policy research that, to our knowledge, has only previously been considered in \cite{papadogeorgou_causal_2017}.  Long-range pollution transport according to atmospheric processes is ubiquitous to the study of pollution interventions at point sources (e.g., power plants or factories), and formal methods for statistical evaluation are lacking for such interventions.  Despite the progress herein, the clustering and partial interference assumption employed in the analysis of SnCR systems is a nontrivial simplification of actual pollution transport, and produces only an approximate analysis of the effect of SnCR on Medicare CVD hospitalizations.  Extensions to more general interference patterns are essential, and a topic of ongoing work.

\section*{Acknowledgments}
The authors thank Dr. Chanmin Kim for helpful discussions in preliminary graphical illustrations and Dr. Christine Choirat for operationalizing the assignment of zip codes to clusters of power plants.   This work was supported by research funding from NIH R01ES026217 and EPA 83587201. Its contents are solely the responsibility of the grantee and do not necessarily represent the official views of the USEPA. Further, USEPA does not endorse the purchase of any commercial products or services mentioned in the publication.

\bibliographystyle{plain}
\bibliography{PowerPlantsR01}

\end{document}